\documentclass[preprint1]{aastex631}
\usepackage{hyperref}
\hypersetup{
    colorlinks=true,
    linkcolor=blue,
    filecolor=magenta,      
    urlcolor=cyan,
    }

\shorttitle{A 4-Gyr M Dwarf Gyrochrone}
\shortauthors{Dungee et al.}

\graphicspath{{./}{figures/}}

\begin{document}

\title{A 4-Gyr M Dwarf Gyrochrone from CFHT/MegaPrime
Monitoring of the Open Cluster M67}


\author[0000-0001-6669-0217]{Ryan Dungee}
\affiliation{Institute for Astronomy, University of Hawai'i, 40 North A`oh\={o}k\={u} Place, Hilo, HI, 96720, USA}

\author[0000-0002-4284-8638]{Jennifer van Saders}
\affiliation{Institute for Astronomy, University of Hawai'i, 2680 Woodlawn Drive, Honolulu, HI, 96822, USA}

\author[0000-0002-5258-6846]{Eric Gaidos}
\affiliation{Department of Earth Sciences, University of Hawai'i at M\={a}noa, 1680 East-West Road, Honolulu, HI, 96822, USA}

\author[0000-0002-8462-0703]{Mark Chun}
\affiliation{Institute for Astronomy, University of Hawai'i, 40 North A`oh\={o}k\={u} Place, Hilo, HI, 96720, USA}

\author[0000-0002-8854-3776]{Rafael A. Garc\'\i a}
\affiliation{AIM, CEA, CNRS, Universit\'e Paris-Saclay, Universit\'e de Paris, Sorbonne Paris Cit\'e, F-91191 Gif-sur-Yvette, France}

\author[0000-0002-7965-2815]{Eugene A. Magnier}
\affiliation{Institute for Astronomy, University of Hawai'i, 2680 Woodlawn Drive, Honolulu, HI, 96822, USA}

\author[0000-0002-0129-0316]{Savita Mathur}
\affiliation{Instituto de Astrof\'isica de Canarias (IAC), E-38205 La Laguna, Tenerife, Spain}
\affiliation{Universidad de La Laguna (ULL), Departamento de Astrof\'isica, E-38206 La Laguna, Tenerife, Spain}

\author[0000-0001-7195-6542]{\^Angela R. G. Santos}
\affiliation{Instituto de Astrof\'isica e Ci\^encias do Espa\c{c}o, Universidade do Porto, CAUP, Rua das Estrelas, PT4150-762 Porto, Portugal}

\correspondingauthor{Ryan Dungee}
\email{rdungee@hawaii.edu}

\begin{abstract}

We present stellar rotation periods for late K and early M dwarf members of the 4 Gyr-old open cluster M67 as calibrators for gyrochronology and tests of stellar spin-down models.
Using \textit{Gaia} EDR3 astrometry for cluster membership, and Pan-STARRS (PS1) photometry for binary identification, we build this set of rotation periods from a campaign of monitoring M67 with the Canada France Hawaii Telescope's MegaPrime wide field imager.
We identify 1807 members of M67, of which 294 are candidate single members with significant rotation period detections.
Moreover, we fit a polynomial to the period versus color-derived effective temperature sequence observed in our data.
We find that the rotation of very cool dwarfs can be explained by simple solid body spin down between 2.7 and 4 Gyr.
We compare this rotational sequence to the predictions of gyrochronological models and find that the best match is Skumanich-like spin-down, $P_\mathrm{rot}\propto t^{0.62}$, applied to the sequence of Ruprecht 147.
This suggests that, for spectral types K7 to M0 with near-solar metallicity, once a star resumes spinning down, a simple Skumanich-like relation is sufficient to describe their rotation evolution, at least through the age of M67.
Additionally, for stars in the range M1--M3, our data show that spin-down must have resumed prior to the age of M67, in conflict with the predictions of the latest spin-down models.

\end{abstract}

\keywords{Open star clusters (1160) --- Stellar evolution (1599) --- Stellar ages (1581) --- Stellar rotation (1629)}

\section{Introduction} \label{sec:intro}

A critical piece of understanding the evolution of any system---be it stars, planets, or the Milky Way galaxy itself, is understanding both the order in which events occur as well as their timescales.
To do this properly one requires precise, reliable ages for the stars involved.
M dwarfs are the most numerous stars in the galaxy \citep{gould1996,bochanski2010}, and have higher occurrence rates of small planets compared to higher mass stars \citep{dressing2015,hardegree-ullman2019}.
They also do not fuse heavy elements, and many tens of Gyr must pass before they show perceptible signs of evolution on a Hertzsprung-Russell diagram \citep{Laughlin1997}.
As a result, M dwarfs can serve as particularly excellent tracers of galactic chemical evolution.

However, M dwarfs are also resistant to most methods commonly used for measuring a star's age.
Their evolution on the main sequence is undetectable \citep{Laughlin1997}, there are no observable asteroseismic oscillations \citep{chaplin2011,Berdinas2017,mathur2019}, and their deep convective envelopes burn Li within the first $\sim50$ Myrs \citep{Bildsten1997}.
The age-peculiar velocity relation is only statistical, making it unreliable for individual stars, and it breaks down for stars that are too young or too old \citep{Aumer2009,lu2021}.
The age-metallicity relationship for the Milky way has flattened out over the past few Gyrs \citep{Holmberg2007}.

Fortunately, rotation period-age relations, or \textit{gyrochronology}, show promise for M dwarf age-dating \citep{barnes2003}.
Gyrochronology relies on the fact that a star spins down over time due to the interaction of its magnetic field with stellar winds, causing a loss of angular momentum \citep{weber1967,Skumanich1972,Barnes2007}.
Observations of Sun-like stars have shown that this angular momentum loss rate, $\frac{dJ}{dt}$, scales strongly with the angular rotation velocity ($\omega$) of the star, $\frac{dJ}{dt} \propto \omega^3$ \citep{Skumanich1972,kawaler1988,Mamajek2008,Meibom2009,Angus2015,gallet2015}, and as a result a star's initial rotation period ($P_\mathrm{rot}$) becomes less important with age \citep{epstein2014,gallet2015}.
The availability of independent age-dating techniques for Sun-like stars such as asteroseismology and isochrone analysis has facilitated the calibration of this age-rotation relationship.
This can be extended to M dwarfs but requires coeval populations of stars (i.e. clusters) of established ages in which the rotation periods of M dwarfs can be obtained.

The rotation period of a star as it hits the main sequence largely depends on two factors: the rotation period the star was born with ($P_\mathrm{init}$) and the protostellar disk lifetime.
Using observations of the youngest pre-main-sequence clusters to fix the values of $P_\mathrm{init}$ indicates that slower rotators are likely to originate from longer disk lifetimes \citep{somers2017,roquette2021}.
The diversity of $P_\mathrm{init}$ values and disk lifetimes leads to a spread of rotation periods ($\sim0.2-8$ days) at the zero age main sequence.
Regardless of the rotation period a star has once it reaches the main sequence, the evolution is dominated by spin down.
As these stars forget their initial conditions they converge onto the slow rotator sequence, a well-defined sequence in temperature-period space.
Prior to this the presence of any stars on a fast rotator sequence or those still heavily-influenced by their $P_\mathrm{init}$ value will make any rotation-age relation ambiguous.
Since accurate gyrochronology relies on a star's initial rotation period becoming less important over time, it relies on convergence onto the slow rotator sequence.

One of the mechanisms by which these fast rotators delay their convergence is saturated spin down.
Saturated spin-down occurs for stars with rotation rates greater than a critical value \citep[$\omega_\mathrm{crit}$, a function of mass;][]{epstein2014}, where spin down scales as $\frac{dJ}{dt} \propto \omega\cdot\omega_\mathrm{crit}^2$ \citep{krishnamurthi1997}.
This is driven by a saturation of magnetic activity, which can be quantified using the  Rossby number ($\mathrm{Ro}$) defined as $(\omega\tau_\mathrm{cz})^{-1}$, where $\tau_\mathrm{cz}$ is the convective overturn timescale.
Generally, smaller Rossby numbers indicate that a star is more magnetically active.
However, below a certain value ($\mathrm{Ro}\lesssim0.1$) stars appear to reach a maximal amount of activity, where decreasing values of $\mathrm{Ro}$ no longer correspond to increases in magnetic activity indicators \citep[][and references therein]{wright2011,matt2015}.
This indicates that particularly fast rotators undergo a decoupling of their rotation rate and their magnetic field strength, resulting in a weaker scaling of torque with rotation rate.
Because $\tau_\mathrm{cz}$ increases with mass, M dwarfs can remain in the saturated regime longer than their higher mass counterparts, which can be seen in the high number of M dwarf fast rotators in clusters such as Praesepe \citep[$670\pm67$ Myr][]{douglas2017,rebull2017} and the Hyades \citep[$728\pm71$ Myr][]{douglas2019}.
Knowing when these fast rotators finally converge is critical for M dwarf gyrochronology.

Observations of solar-mass stars younger than the Hyades ($\lesssim600$ Myr) have shown that models which assume the entire star rotates with a uniform angular velocity (i.e., solid body rotation) fail to match the observed convergence onto a slow rotator sequence and subsequent evolution \citep{keppens1995,krishnamurthi1997,allain1998}.
Models which incorporate the internal transport of angular momentum (i.e., differential rotation) relax the assumption of solid body rotation.
In particular, core-envelope decoupling models take a simplified approach of treating the core and envelope as two separate, rotationally solid bodies with a mechanism that transports angular momentum between the two \citep{macgregor1991,denissenkov2010,lanzafame2015}.
The critical parameter, then, is the timescale over which torques act to equilibriate the rotation rates ($\tau_\mathrm{c-e}$).
Fits to cluster data have shown $\tau_\mathrm{c-e}$ is $\sim20$ Myr for solar-mass stars and a strong function of mass \citep[$\tau_\mathrm{c-e}\propto M^{-7.28}$ or $M^{-9.1\pm1.8}$ by][respectively]{lanzafame2015,somers2016}.

Recent observations of a collection of open clusters, namely Praesepe \citep{douglas2017,rebull2017}, the Hyades \citep{douglas2019}, NGC 6811 \citep{curtis2019}, and NGC 752 \citep{agueros2018}, have shown that K and early M dwarfs appear to halt their spin down for a period of time---a striking departure from a standard spin-down model with solid body rotation, but a phenomenon that can be explained by core-envelope decoupling models \citep{spada2020}.
In \citet{spada2020} the apparent stalling is caused by the angular momentum loss of the envelope being balanced by transport from the core.
The result is a net loss of angular momentum from the star, while the envelope continues to rotate at a roughly constant rate.
However, their model does not predict the same degree of stalling as observed in open clusters, as it predicts that stars later than K5 should be rotating $\sim5$ days slower than they are in Ruprecht 147 \citep{curtis2020}.
We know that K and M dwarfs must resume spinning down, as field samples show K and M dwarfs with rotation periods that are many tens of days \citep{mcquillan2013,newton2016,newton2017,santos2019}. Such rotation periods are roughly consistent with a Skumanich-type spin-down over the age of the galactic disk \citep{vanSaders2019}.
Knowing when these stars resume spinning down and the timescales over which internal angular momentum exchange occur both directly affect the mapping of a rotation period to an age.

Calibrating gyrochronology and testing spin-down models for M dwarfs requires a larger sample of older, well-dated M dwarfs.
Only a handful of such stars are currently available, the majority of which are in young clusters \citep[700 Myr at the oldest;][]{douglas2017,rebull2017} or are limited by the use of kinematic ages \citep{newton2016,popinchalk2021}.
Previously, \citet{barnes2016} used \textit{K2} to obtain calibrators for gyrochronology of solar-type stars in M67, but observing faint M dwarfs in crowded fields has proved impossible for missions such as \textit{K2} or \textit{TESS}.
The Canada France Hawaii Telescope's MegaPrime \citep{boulade2003} instrument allows us to overcome the limitations of \textit{K2} and \textit{TESS} in sensitivity, without significant loss in field-of-view.
In this paper we present the rotation periods of late K and early M dwarf members of the 4 Gyr-old cluster M67 \citep[3.5-5.0 Gyr;][]{nissen1987,demarque1992,montgomery1993,carraro1994,fan1996,vandenberg2004,balaguernunez2007,stello2016}.
We present the oldest K and M dwarf gyrochrone to date and compare it to literature gyrochronology relations.

\section{Observations and Data Reduction} \label{sec:obs&reduce}

\begin{figure}
    \centering
    \includegraphics[scale=0.6]{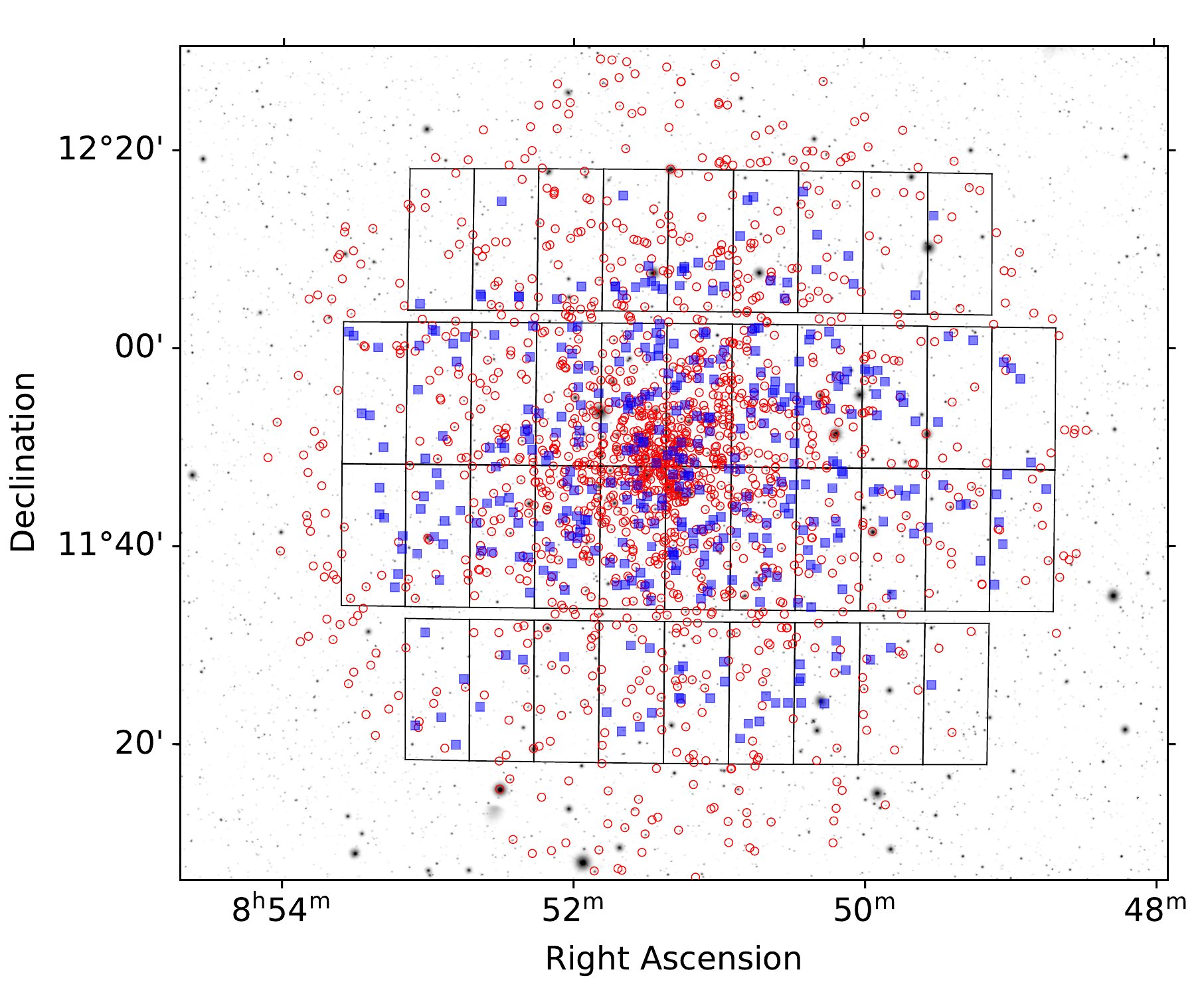}
    \caption{A Sloan Digital Sky Survey DR9 $i$-band image of the M67 field, with all of our candidate M67 cluster members identified by circles. Blue circles identify the the stars with reported rotation periods in Table \ref{tab:results}. The black boxes are the MegaPrime footprint for one of the pointings in our dither pattern.}
    \label{fig:field_image}
\end{figure}

Our campaign used Canada France Hawaii Telescope's (CFHT) MegaPrime to monitor M67 (center coordinates: $\alpha=8^\mathrm{h} 51^\mathrm{m} 18^\mathrm{s}$, $\delta=+11^\circ48'00''$) from 2018 October 15 to 2021 March 5 (UT).
MegaPrime is the MegaCam imager placed at the prime focus of CFHT; it has a 1 square degree field-of-view sampled by 40 CCDs arranged in 4 rows of 9, 11, 11, and 9 detectors each (Fig.\ \ref{fig:field_image}).
In total we obtained 694 exposures of the cluster in discrete 1-2 week runs, producing 131 epochs of data for our light curves.
All data were collected with 121 second integration times using the Sloan $i$ filter, red enough that M dwarfs are not too faint but blue enough to observe spot variability.
Observations were taken five at a time in a cross-like dither pattern with 10.4 arcsec offsets.
Bias subtracting, flat-fielding, fringe correction, and bad-pixel masking were all performed by version 3.0 of the CFHT \texttt{Elixir} pipeline \citep{magnier2004}.
For our data reduction we treated the five exposures in a dither pattern independently, only combining them when we averaged the five photometric measurements together in the later steps of the pipeline.
An image of the field taken from $i$-band images of the Sloan Digital Sky Survey DR9 is shown in Fig.\ \ref{fig:field_image}.
Included are the candidate cluster members of M67 (see Sec.\ \ref{sec:membership} for further details).

\subsection{Sky Background} \label{subsec:skybkg}

We started with determining and removing the sky background from each detector's image.
For this we used \texttt{MMMBackground}, a \texttt{python} implementation of the DAOPHOT MMM algorithm contained in the \texttt{photutils} package \citep{stetson1987,photutils}.
We divided each image into a 8x10 grid of equally sized sub-regions and for each sub-region we estimated the background level through an estimation of the mode by the equation $Mode \approx 3 \times Median - 2 \times Mean$.
This 8x10 grid was then interpolated to the size of the original image using a bi-cubic spline and the resulting sky background was subtracted from the image.
To estimate the uncertainty on this sky background we also computed the sigma-clipped standard deviation of each sub-region in the grid which was similarly interpolated to produce an estimated uncertainty for each pixel in the sky background.

\subsection{Source Finding}\label{subsec:sourcefinding}

With the sky background subtracted, we then used \texttt{DAOStarFinder}, a \texttt{python} implementation of the DAOFIND algorithm in the package \texttt{photutils}, to find the location of every source in the field for each individual image \citep{stetson1987,photutils}.
The threshold was set relatively low, at three times the sigma-clipped standard deviation of all pixel values in the image, and the full width at half maximum (FWHM) was set at the seeing value reported in the image header.
Using the World Coordinate System values in the image headers, we converted the pixel coordinates reported by \texttt{DAOStarFinder} to RA and Dec (J2000.0).
This enabled us to cross-match our detected sources with external catalogs.

We downloaded a catalog of every \textit{Gaia} EDR3 \citep{gaiaedr3} source in the field of view and converted the RA and Dec coordinates to the epoch J2000.0.
For each \textit{Gaia} source we then found the nearest neighbor match reported by \texttt{DAOStarFinder}.
A handful of \textit{Gaia} sources did not have a detection within the cut-off of 0.75 arcsec and were considered non-detections.
Any remaining sources found by \texttt{DAOStarFinder} that were not paired up with a \textit{Gaia} source were considered false positives and discarded from our catalog.
A nearest neighbor search, with the same distance cut-off, was also used to match every \textit{Gaia} source to a Pan-STARRS1 (PS1) DR2 source.
After this cross-matching our catalog contained 8287 sources, all of which were matched to a source in both \textit{Gaia} EDR3 and PS1 DR2.
The limiting magnitudes of our observations ($i\sim21$) compared to that of \textit{Gaia} EDR3 ($G\sim21$) and PS1 DR2 ($i\sim22$) limit the number of real sources that were discarded by this method.

\subsection{Photometry}\label{subsec:phot}

Next we performed aperture photometry on the background-subtracted images for every source in our catalog.
The aperture diameter was set at four times the seeing value for the image, a value computed from the average empirical full width at half maximum of bright sources scattered throughout the field.
This diameter was chosen after analyzing the effect of aperture size on the noise properties of the light curves (see related discussion in Sec.\ \ref{subsec:valid}).
Any sources with overlapping apertures were flagged and excluded from the calculation of the zero-point corrections described in this section due to the source confusion introduced by their overlap.
Instrumental magnitudes were computed by a sum over the aperture divided by exposure time, and an uncertainty was estimated from the quadrature sum of: the photon noise on the flux in the aperture, the read noise of the MegaPrime detectors, and the previously estimated sky uncertainty for each pixel in the aperture.
With this process repeated for each of our observations, we then began constructing the light curves for each target.

In order to construct the light curves we first averaged together the measurements within an epoch (i.e., the set of five exposures that make up one dither pattern).
To do this we corrected for small changes in the photometric zero-point that may have occurred between exposures.
The correction, which we call $\Delta_i$, was taken to be the median difference between the stars with low scatter in their instrumental magnitudes.
In equation form, the correction applied to the $i$th observation relative to the first ($i=0$) is given by:
\begin{equation}
\Delta_{i} = \mathrm{Median}(M_{ls, 0} - M_{ls, i})
\end{equation}
Where $M_{ls, i}$ represents the set of stars with less than median scatter in their $n$ measurements.
Since any measurement where the aperture included bad pixels is discarded, $n \leq 5$.
The epoch magnitude was then computed from the average of the zero point-corrected measurements.
If $m_i$ is the instrumental magnitude from the $i$th observation within the $j$th epoch then $\bar{m}_j$, which will become a point in a light curve, is given by:
\begin{equation}
\bar{m}_j  = \frac{1}{n} \sum_{i=0}^{n-1} (m_i + \Delta_{i})
\end{equation}
We repeated this calculation for every star and every epoch.

Finally, we corrected for the changes in the photometric zero-point between epochs.
We took the same approach as before, computing this correction from the low scatter stars.
The zero-point correction, $\mathrm{zp}_j$, of the $j$th epoch relative to the first ($j=0)$ is given by:
\begin{equation}
\mathrm{zp}_{j} = \mathrm{Median}(\bar{M}_{ls, 0} - \bar{M}_{ls, j})
\end{equation}
Where $\bar{M}_{ls, j}$ represents the set of stars with less than median scatter in their computed $\bar{m}$ values.
Each $\bar{m}$ was then corrected by this zero-point correction:
\begin{equation}
m_{\mathrm{epoch},j} = \bar{m}_j - \mathrm{zp}_j
\end{equation}
In total, we obtained 4396 light curves that met our completeness criterion of at least 99 epochs of available data (see Sec. \ref{sec:rotation} for further details).

\subsection{Validation} \label{subsec:valid}

\begin{figure}
    \centering
    \includegraphics[scale=0.57]{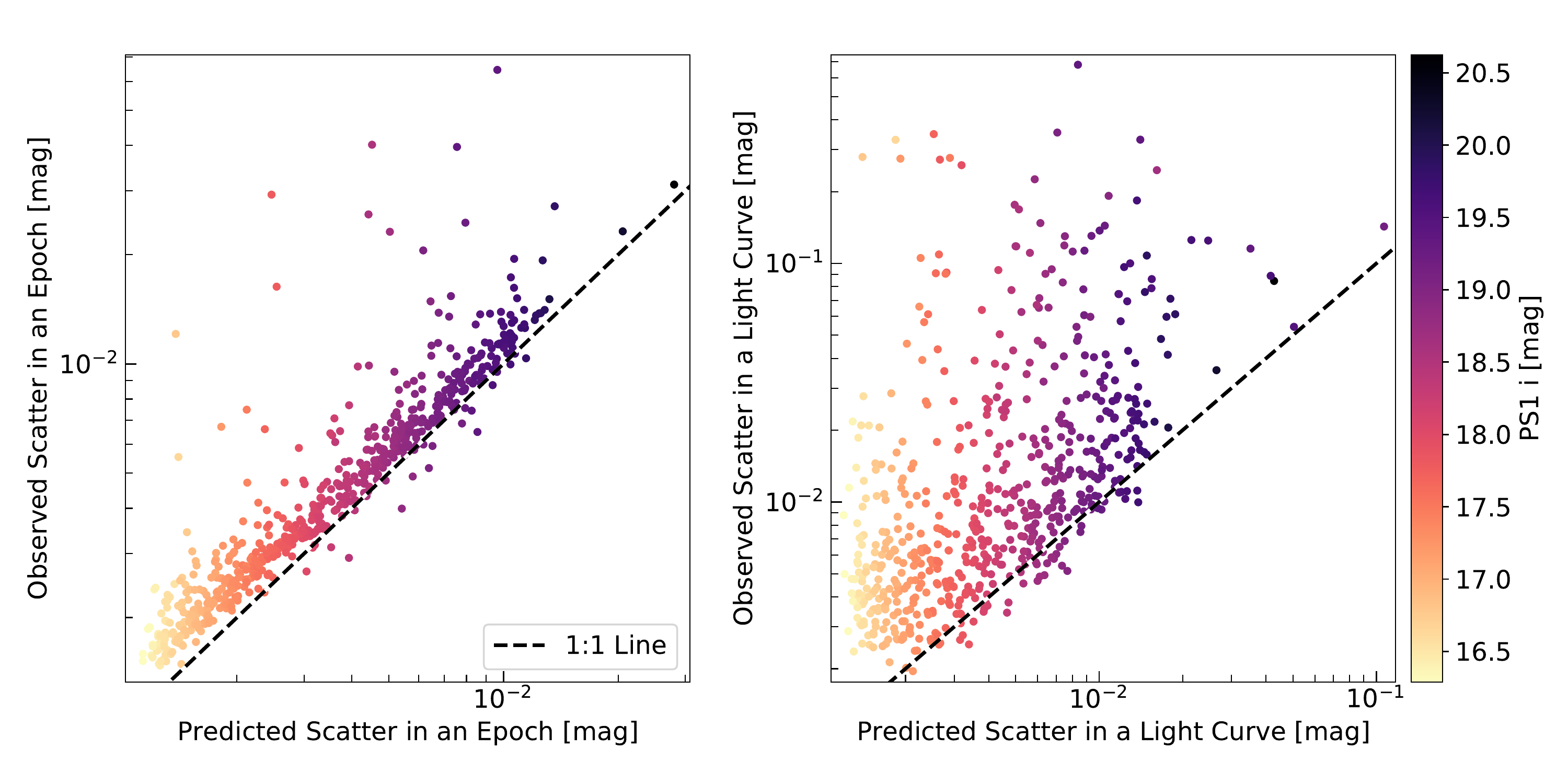}
    \caption{\textbf{Left Panel}: A scatter plot of the predicted versus observed scatter of the zero-point corrected magnitudes that are averaged together to form one epoch in our light curves. The expectation is that these points lie on the one-to-one line. \textbf{Right Panel}: A scatter plot of the predicted versus observed scatter of the zero-point corrected magnitudes in our light curves, the one-to-one line is the expected lower-limit. In both panels the predicted values are derived from our estimated uncertainties.}
    \label{fig:sidm_mids}
\end{figure}

In order to validate our model of the photometric noise we performed a comparison of the observed scatter in the zero-point corrected magnitudes to the scatter expected from the estimated uncertainties alone.
We made two important assumptions for these tests: 1) the uncertainties were the standard deviation of independent Gaussian distributions, and 2) all of these distributions had the same mean (i.e., the uncertain measurement was the only source of variability).
Thus, any sources with additional variability in their magnitudes would fall above the one-to-one line on a plot of the theoretical scatter versus the observed scatter.

First, we performed this test on the $n\leq5$ measurements that form an epoch, comparing the standard deviation of these points to the scatter expected from the uncertainty on their average.
Exposures in an epoch were collected over a period of roughly 20 minutes, short enough that we expected each star not to vary.
As a result, a scatter plot of how the noise was modeled versus the observed scatter should follow a one-to-one line, as we see in our data (left panel of Fig. \ref{fig:sidm_mids}).
For sources with magnitudes of $i\lesssim18$, the data show a departure from the one-to-one line which we attribute to either non-linearity in the detector as it approaches saturation or a fractional measurement error, such as flat fielding errors.

Second, we performed this test on full light curves, comparing the standard deviation of the epoch magnitudes to the scatter expected from their uncertainties.
The one-to-one line is expected to be the lower limit of photometric scatter, therefore many sources will show variability beyond the case of random variations due to uncertain measurements of the magnitude.
We stress that a source falling above the one-to-one line in the right panel of Fig.\ \ref{fig:sidm_mids}, meaning it has more variability than expected from uncertainty in the photometry alone, is not proof of the source having an astrophysical process driving that variability.
That the one-to-one line is indeed the lower limit in the right panel of Fig. \ref{fig:sidm_mids} demonstrates that our model of the noise is correct.

\section{Cluster Membership and Stellar Properties} \label{sec:membership}

An important aspect of the results presented in this paper is that the rotation periods reported can be used as benchmarks for stellar spin down models. Critical to this is knowing the age and $T_\mathrm{eff}$.
The age determination comes from their membership in the open cluster M67, whose age has been previously determined to be 4 Gyr \citep[3.5--5 Gyr;][]{nissen1987,demarque1992,montgomery1993,carraro1994,fan1996,vandenberg2004,balaguernunez2007,stello2016}. The effective temperatures are derived using a color-$T_{\mathrm{eff}}$ relation. In this section, we provide the details on both of these critical aspects.

\subsection{Cluster Membership} \label{subsec:membership}

\begin{figure}
    \centering
    \includegraphics[scale=0.7]{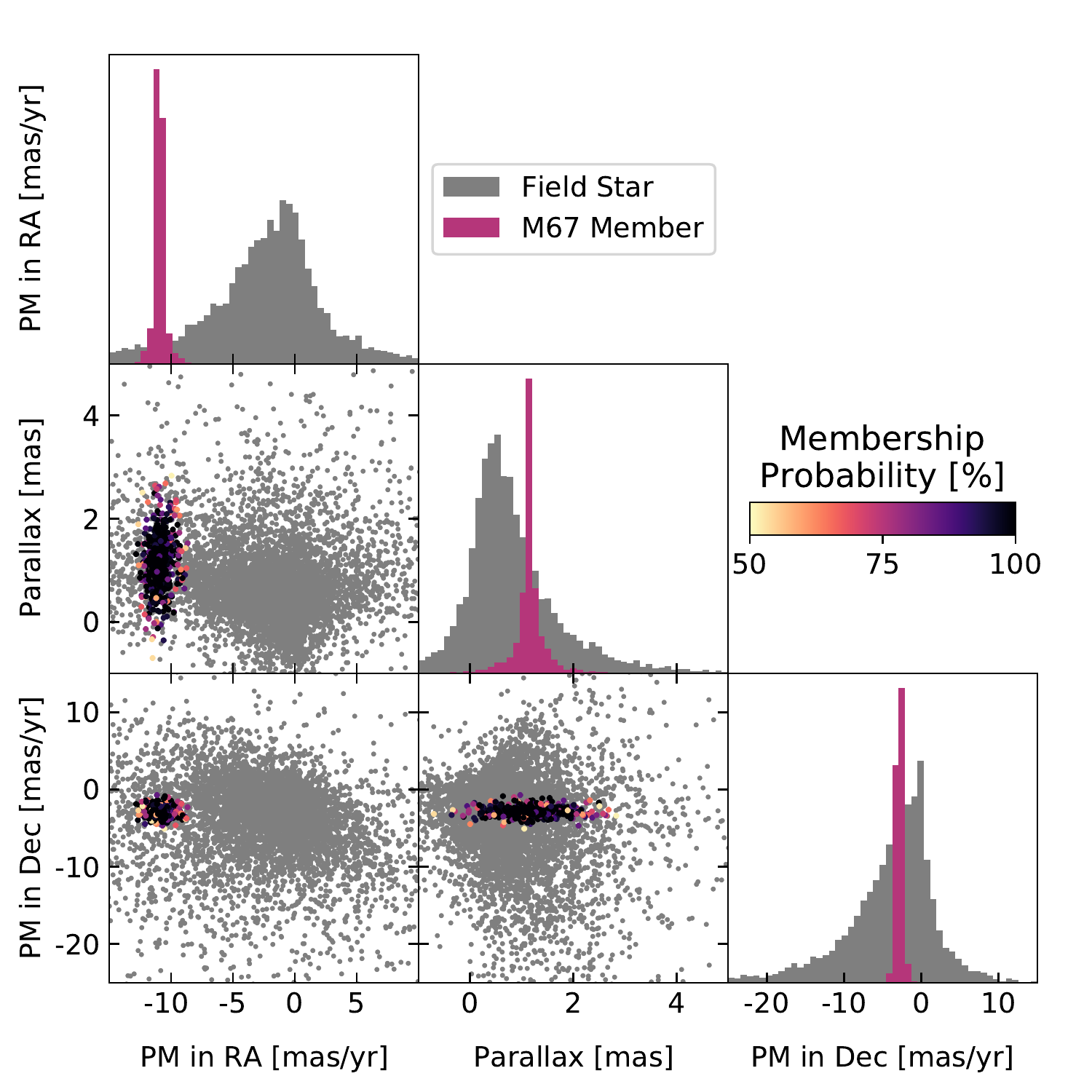}
    \caption{Scatter plots and histograms of the three parameters used for cluster membership determination (proper motions, or PM, in RA and Dec, and parallaxes). In gray, the distribution of these parameters for the non-cluster stars as a point of comparison. For cluster members color indicates the probability of membership, showing that the members on the ``outskirts" of the distribution are less likely to be considered members of M67 (see Sec.\ \ref{subsec:membership} for further details).}
    \label{fig:kine_scatter}
\end{figure}

Cluster membership was determined by using a clustering algorithm, \texttt{HDBSCAN} \citep{hdbscan}, on \textit{Gaia} EDR3 proper motions and parallaxes for every star in the MegaPrime field of view, regardless of whether or not it appeared in our catalog of light curves.
\texttt{HDBSCAN} works by using the density of points to estimate a probability distribution function (PDF) that describes the full distribution of values in the data.
Clusters are then defined by the peaks in this PDF.
The primary advantage of \texttt{HDBSCAN} is that it relies on fewer assumptions about the data than more traditional clustering algorithms such as K-means, which assumes Gaussian distributions.
It also does not require that every point in the data set be assigned to a cluster, reducing the risk that outlier field stars might incorrectly be assigned M67 membership.
The \texttt{python} package of the same name provides many different parameters to tune the performance of the clustering\footnote{Details on the parameters can be found under Parameter Selection for HDBSCAN* \href{https://hdbscan.readthedocs.io/en/0.8.18/}{in the docs}}..
We found that the defaults for the version we used, v0.8.18, were acceptable with one exception: \texttt{min\_samples}.
This parameter can be thought of as determining the level of detail in \texttt{HDBSCAN}'s estimation of the underlying PDF.
Too small a value and each data point produces its own peak in the estimated PDF, too large and the finer details of the estimated PDF are washed out.
Given that the distribution of parallaxes and proper motions is effectively a two-peaked distribution (Fig.\ \ref{fig:kine_scatter}) we found that a value of 200 gave suitable clustering results compared to the default of 5.

A disadvantage of the \texttt{HDBSCAN} algorithm is that it does not make use of the uncertainties on any input data.
To incorporate these into our cluster membership determination we performed Monte Carlo sampling.
We ran the clustering algorithm on our list of \textit{Gaia} sources, recorded the results, and resampled every star assuming Gaussian uncertainties, repeating this process 1000 times.
\texttt{HDBSCAN} cannot be instructed to find a cluster with specific properties, so with each realization we computed the median parallax and proper motions for each grouping it found in the data and used the one with the closest match to literature values for M67 \citep[$\pi=1.1327\pm0.0018$ mas, $\mu_\alpha\cos \delta=-10.9738\pm0.0078$, $\mu_\delta=-2.9465\pm0.0074$ mas yr$^{-1}$;][]{gao2018}.
The difference was never more than a few percent.
We then calculated a ``kinematic membership probability" from the fraction of realizations in which a star was assigned membership to M67.
A final membership criterion of $>50\%$ was selected based on an inspection of the \textit{Gaia} color-magnitude diagrams (CMDs, Fig.\ \ref{fig:cmd}) that were produced for various thresholds.

Another aspect we considered in our use of \texttt{HDBSCAN} was the Bayseian nature of this approach.
Too small a field of view and the algorithm may not have had the leverage needed to separate M67 members from the field, too large and the diversity of field stars may have encouraged labelling true members as field stars.
To address this we repeated our membership determination on a \textit{Gaia} EDR3 catalog including stars out to twice the radius of the MegaPrime field of view and compared the two membership lists.
Of the 1807 members within the MegaPrime field of view 76 were considered field stars when using the larger catalog, and none of the field stars gained membership in M67.
None of these 76 stars are outliers on our CMDs (Fig.\ \ref{fig:cmd}), so we have kept them in our final list of members.
However, we flagged them as potentially suspect, so that the interested reader may remove them from the sample if they wish.

We compared our list of M67 members to that of \citet{gao2018}, who applied a Gaussian mixture model based approach to \textit{Gaia} DR2 astrometry.
They found a list of 1502 likely members, whereas we have found 1807.
In common between the two catalogs are 1241 members, leaving 261 stars that are unique to the \citet{gao2018} catalog, and 566 that are unique to our catalog.
There are several factors that contributed to these differences.
First, our search for members was limited to the field of view of CFHT MegaPrime, and this truncated our search at a radius of $\sim$30 arcminutes; all 261 stars that are only in the \citet{gao2018} catalog were outside our field of view and thus were not included in our clustering.
Second, 431 of the 566 stars that appear only in our catalog have parallaxes and/or proper motions that are closer to the literature values for M67 in EDR3 than in DR2.
Third, 99 of the 566 stars that appear in our catalog are new in EDR3 and thus could not have been included in the \citet{gao2018} catalog.
Finally, there are 36 stars which are unique to our catalog for otherwise unknown reasons, we attribute these to the differences between the two methods used.
The members that our two catalogs have in common are denoted by the ``Gao member" column in Table \ref{tab:results}.

\subsubsection{Single vs Binary Members} \label{subsubsec:binaries}

\begin{figure}
    \centering
    \includegraphics[scale=0.6]{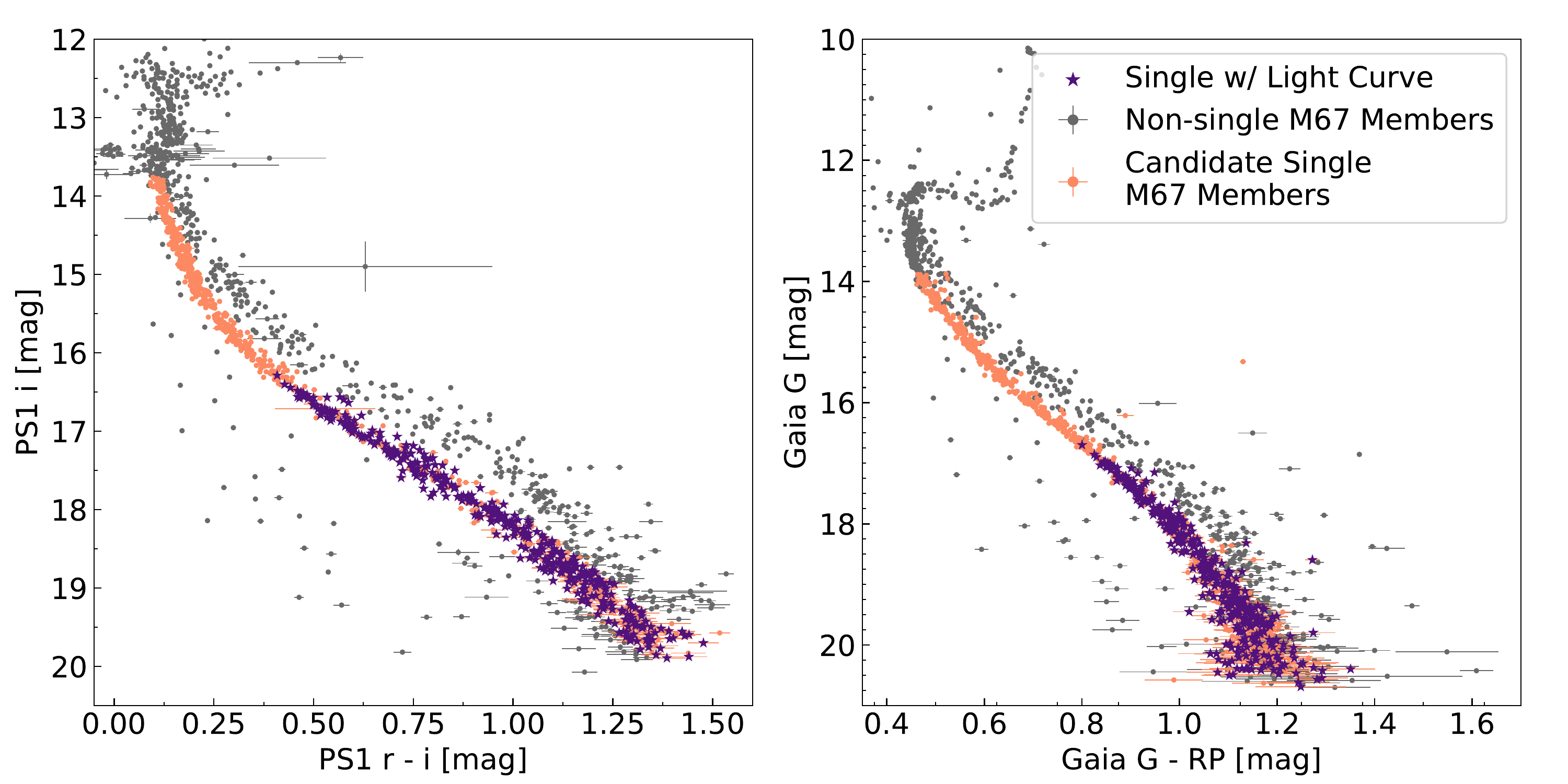}
    \caption{The color magnitude-diagram of kinematically selected M67 members. Left panel is using the Pan-STARRS DR2 photometry and right panel is using \textit{Gaia} EDR3 photometry. In both panels orange points represent main sequence single members (subject to a brightness cutoff at lower magnitudes), grey points represent photometric binaries, and purple points indicate stars with rotation periods reported in our results.}
    \label{fig:cmd}
\end{figure}

\begin{figure}
    \centering
    \includegraphics[scale=0.6]{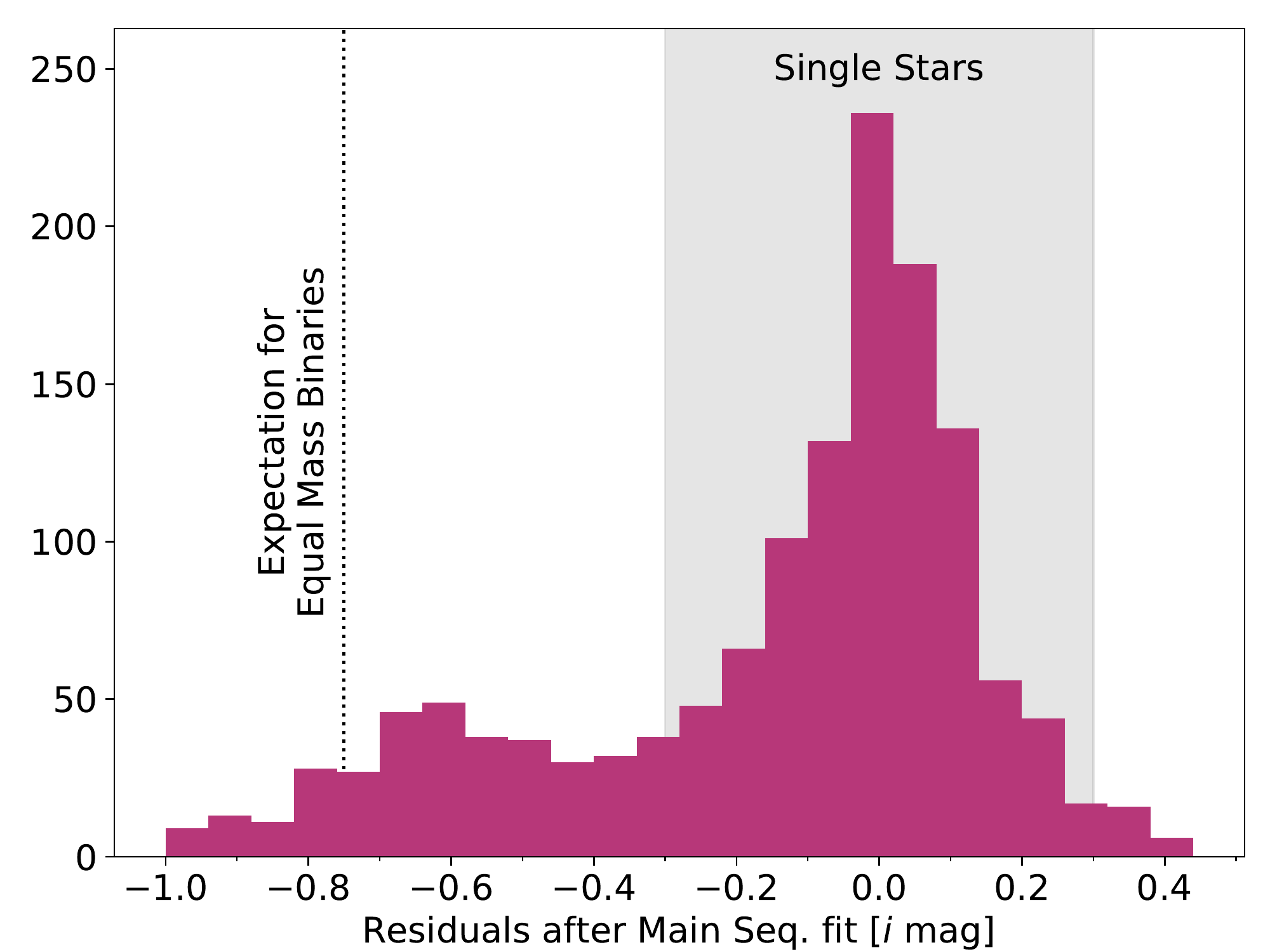}
    \caption{The distribution of residuals after subtracting out a polynomial fit to the main sequence from each star in the catalog. The shaded region denotes the stars chosen as single stars on the main sequence. The -0.3 offset represents the approximate location of the valley between the two peaks in this distribution, while accounting for the wide spread about the binary peak centered at $\sim-0.6$. The vertical dotted line denotes the expected excess brightness for equal mass binaries.}
    \label{fig:bin_v_single}
\end{figure}

Unresolved binaries bias our inferred stellar parameters and close binaries have spin-down influenced by tidal forces \citep{simonian2020}; as a result we also need to identify whether or not the M67 members have a companion.
With a parallax of $\pi=1.1327\pm0.0018$ mas, \textit{Gaia} is able to resolve binaries that are separated by $\gtrsim 600$ AU.
However, due to the size of our apertures stars with physical separations $\lesssim7000$ AU have overlapping apertures. As such, their photometry was potentially limited by confusion with their nearest neighbor.
For completeness, we included these stars in our catalog of reported rotation periods (Sec.\ \ref{sec:res}), but we excluded them from our subsequent analysis.
Work done by \citet{deacon2020} indicates there are no wide binaries separated by $\gtrsim3000$ AU in clusters, thus our analysis was focused only on single members of M67.
Binary systems which are not resolved require a different method of detection. Common approaches include: 1) spectroscopy that resolves double-lined absorption features, 2) excess astrometric noise \citep[quantified by the renormalized unit weight error, or RUWE, for \textit{Gaia} astrometric solutions;][]{belokurov2020}, and 3) photometric excess, stars that appear brighter than the main sequence on a CMD.
For our data we used the photometric excess approach, calculated from PS1 photometry \citep{ps1phot,ps1db}.
The effectiveness of this approach was confirmed by the finding that all the sources which exhibited excess astrometric noise (i.e., RUWE $\gtrsim1.4$) were also found to show photometric excess.

In PS1 DR2 the saturation limit is 12-14 magnitudes, depending on seeing and filter; for M67 we found that a cut at $i=13.75$ removed these problematic sources.
From the photometry in \textit{Gaia} EDR3 and PS1 DR2 we found that PS1 $r$ and $i$ were the two filters with the highest signal-to-noise ratios for the faintest members in our catalog.
Therefore, we used $3\sigma$ iterative outlier rejection to fit an eighth order polynomial to the main sequence of the cluster on a PS1 $r-i$ vs $i$ CMD.
We then categorized each source by its vertical distance from the main sequence on the CMD based on the distribution of residuals after subtracting out our fit to the main sequence (Fig.\ \ref{fig:bin_v_single}).
Stars within $\pm0.3$ magnitudes of the main sequence fit were classified as single members, whereas sources outside these bounds were categorized as photometric binaries.
The small secondary peak of binary members is broad, which suggests there may be a small number of high-contrast binaries contaminating our sample of candidate single M67 members.
Sources fainter than the main sequence are thought to be binaries with a white dwarf component.
Further observations are required to confirm this, which will be included in future work.
The fraction of sources labeled as binaries by this method is $26\%$, in line with the expectation for M dwarf multiplicity rates \citep{duchene2013,winters2019}.
Both the PS1 $r-i$ vs $i$ CMD that was used for this binary classification and an additional \textit{Gaia} $G-RP$ vs $G$ CMD can be seen in Fig.\ \ref{fig:cmd}.
The stars identified as binaries by this method were set aside for future analysis. They do not have rotation periods reported in this paper.

\subsection{Effective Temperatures} \label{subsec:teffs}

\begin{deluxetable}{llcl}
\tablecaption{M Dwarfs Used to Derive $T_\mathrm{eff}(r-i)$ \label{tab:teffs}}
\tablehead{\colhead{Name} & \colhead{Table Header} & \colhead{Units} & \colhead{Description}}
\startdata
Object Name & \texttt{name} & -- & Source name used in \citet{mann2015} \\ 
Right Ascension & \texttt{raDeg} & $^\circ$ & -- \\ 
Declination & \texttt{deDeg} & $^\circ$ & -- \\ 
\textit{Gaia} G magnitude & \texttt{gaiaGmag} & mag & The synthetic \textit{Gaia} G magnitude \\
\textit{Gaia} BP magnitude & \texttt{gaiaBPmag} & mag & The synthetic PS1 \textit{Gaia} BP magnitude \\
\textit{Gaia} RP magnitude & \texttt{gaiaRPmag} & mag & The synthetic PS1 \textit{Gaia} RP magnitude \\
Mass & \texttt{solMass} & $\mathrm{M}_\odot$ & Mass of the star \\ 
Error on the Mass & \texttt{e\_solMass} & $\mathrm{M}_\odot$ & -- \\
Metallicity & \texttt{[Fe/H]} & -- & Metallicity of the star \\
Error on the Metallicity & \texttt{e\_[Fe/H]} & -- & -- \\
$\mathrm{T_\mathrm{eff}}$ & \texttt{teff} & K & Effective Temperature \\
PS1 g magnitude & \texttt{ps1gmag} & mag & The synthetic PS1 g magnitude \\
Error on PS1 g magnitude & \texttt{e\_ps1gmag} & mag & -- \\
PS1 r magnitude & \texttt{ps1rmag} & mag & The synthetic PS1 r magnitude \\
Error on PS1 r magnitude & \texttt{e\_ps1rmag} & mag & -- \\
PS1 i magnitude & \texttt{ps1imag} & mag & The synthetic PS1 i magnitude \\
Error on PS1 i magnitude & \texttt{e\_ps1imag} & mag & -- \\
PS1 z magnitude & \texttt{ps1zmag} & mag & The synthetic PS1 z magnitude \\
Error on PS1 z magnitude & \texttt{e\_ps1zmag} & mag & -- \\
Ks magnitude & \texttt{Ksmag} & mag & The synthetic Ks magnitude \\
Error on Ks magnitude & \texttt{e\_Ksmag} & mag & -- \\
\enddata
\tablecomments{This table is available in its entirety in machine-readable format online, here we provide a description of each column in the table.}
\end{deluxetable}

\begin{figure}
    \centering
    \includegraphics[scale=0.6]{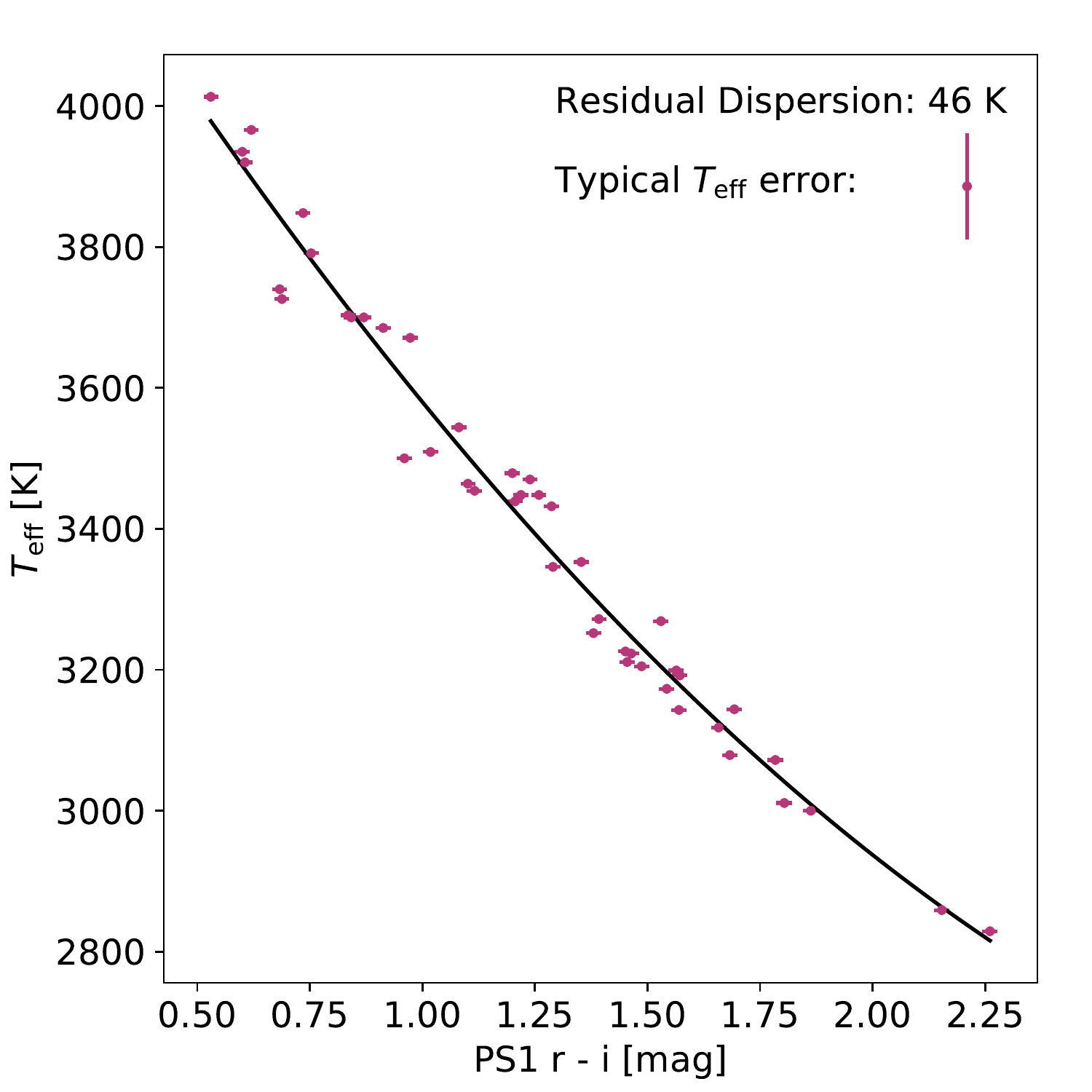}
    \caption{The sample of late K and M dwarfs of \citet{mann2015} used to derive our color-$T_{\mathrm{eff}}$ relationship. We chose stars to have a metallicity within the range of literature values for M67. The $T_\mathrm{eff}$ errors are of order $75\;\mathrm{K}$ and do not affect the dispersion.}
    \label{fig:color2teff}
\end{figure}

In order to calculate the effective temperature ($T_{\mathrm{eff}}$) for the stars in our catalog we used a ($r-i$) vs $T_{\mathrm{eff}}$ relation derived from the sample of late K and M dwarfs analyzed by \citet{mann2015}.
We converted the synthetic Sloan $r$ and $i$ photometry provided into the PS1 $r$ and $i$ passbands using the \citet{tonry2012} relations and applied corrections for reddening \citep[$E(B-V)=0.041\pm0.004$ mag;][]{taylor2007} as well as a conversion to apparent magnitudes for the distance to M67 \citep[$\pi=1.1327\pm0.0018$ mas;][]{gao2018}.
We trimmed the sample to stars with metallicities of $-0.07 \leq [\mathrm{Fe/H}] \leq 0.07$, a range chosen to cover various values reported for the metallicity of M67 in the literature \citep{pace2008,santos2009,onehag2011,liu2016,sandquist2018}.
The parameters of the trimmed sample are included in Table \ref{tab:teffs}.
Finally, we fit a second order polynomial to the ($r-i$)-$T_{\mathrm{eff}}$ pairs to obtain our relation:
\begin{equation}
    \phantom{.} T_{\mathrm{eff}} = 139.8 (r-i)^2 - 1062.1 (r-i) + 4502.1 .
\end{equation}
The residual dispersion of $46$ K is small compared to the errors on the temperatures (Fig.\ \ref{fig:color2teff}).
Adding this in quadrature with the spectroscopic errors provided by \citet{mann2015} yields a $T_{\mathrm{eff}}$ uncertainty of $\sim75$ K.

\section{Measuring Rotation Periods} \label{sec:rotation}

In our data set there are 7222 sources which contain at least one epoch of data.
To reduce complications with recovering periodic signals we applied a conservative cut to our data, requiring that a light curve have a minimum completion of 99 out of the possible 131 epochs of data.
After applying this cut we were left with a sample of 4674 stars, with a mean completeness of 129 epochs.
Of these 4674 stars, 3607 have light curves with 131 epochs.
For the 636 candidate cluster members that made these cuts the mean completeness is 128 epochs, with 444 having a light curve that has 131 epochs.
Due to the irregular sampling of our light curves we used Lomb-Scargle (LS) periodograms \citep{lomb1976,scargle1982,press1989,zechmeister2009} for the detection of periodic signals in our light curves.
In each case the rotation periods we report was the period of maximum power in the periodogram.

A common method for quantifying the uncertainty of LS periodograms is the false alarm probability (FAP).
The FAP is a measure of probability that data with no signal would produce a peak in the periodogram of equivalent height \citep[for further details see Sec.\ 7.4.2 of][]{vanderplas2018}.
We required a FAP value of less than one percent for a periodic signal to be considered significant.
To maintain the computational feasibility of our injection and recovery tests (see Sec.\ \ref{subsec:iandr}) we report FAP values estimated using the \citet{baluev2008} method.
As a test of the validity of using the Baluev estimates, we performed a comparison of the FAP values estimated by the Baluev method to those computed using a bootstrapping ($N=10^4$) of all of our light curves.
Using the bootstrapping method it is roughly expected to find $\mathrm{FAP}\times N \pm \sqrt{\mathrm{FAP} \times N}$ false positives \citep{vanderplas2018}.
Accounting for this uncertainty on the FAP value, every one of our Baluev estimated FAP values is consistent with its bootstrapped equivalent.
We also required that a rotation period have at least five complete periods within the light curve duration in order to be considered a detection.
This placed an upper limit of 175 days on any rotation periods used in our analysis.

\subsection{Injection and Recovery Tests} \label{subsec:iandr}

\begin{figure}
    \centering
    \includegraphics[scale=0.6]{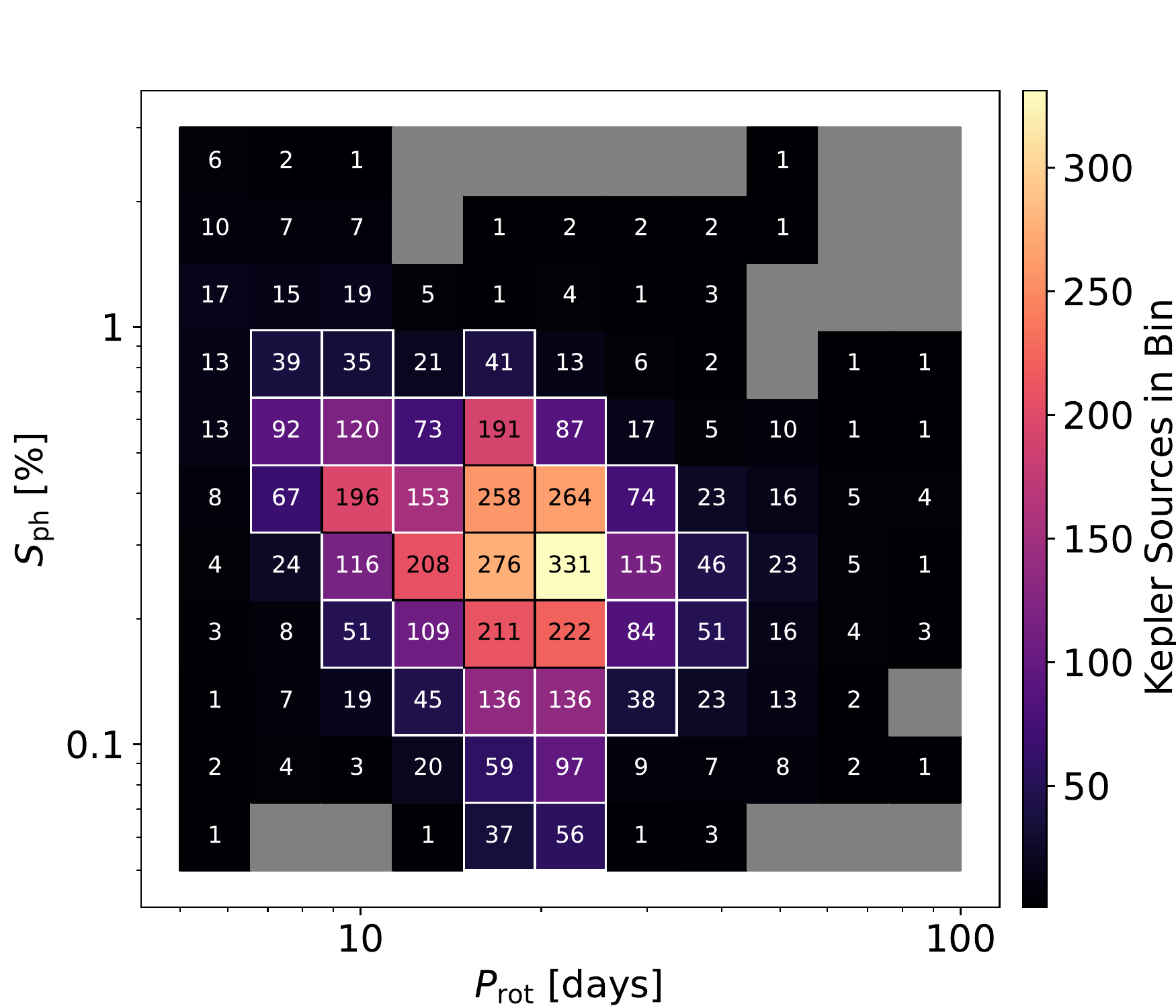}
    \caption{The number of \textit{Kepler} light curves used in the injection and recovery tests to establish completeness as a function of rotation period ($P_\mathrm{rot}$) and photometric variability ($S_\mathrm{ph}$). Bins with a black or white (color chosen for optimal contrast) box drawn around them are the ones where over half (i.e.\ $\geq25$) of the injected signals are \textit{Kepler} light curves.}
    \label{fig:kplr_in_bins}
\end{figure}

\begin{figure}
    \centering
    \includegraphics[scale=0.6]{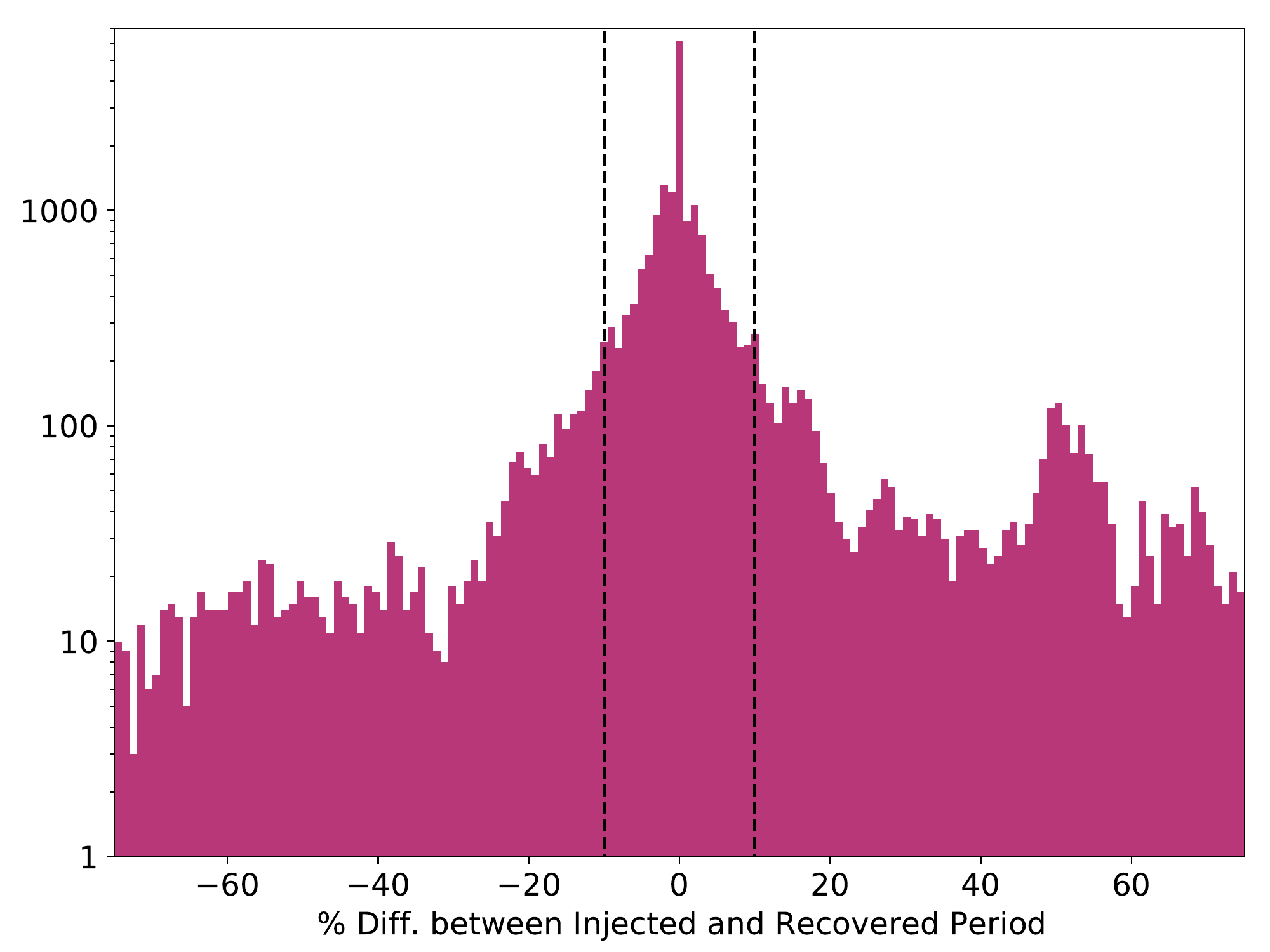}
    \caption{The distribution of the percent differences between the injected and recovered period (irrespective of the false alarm probability of the recovery). The precision on our rotation periods is set by the standard deviation of this distribution: $10\%$.}
    \label{fig:ls_precision}
\end{figure}

\begin{figure}
    \centering
    \includegraphics[scale=0.6]{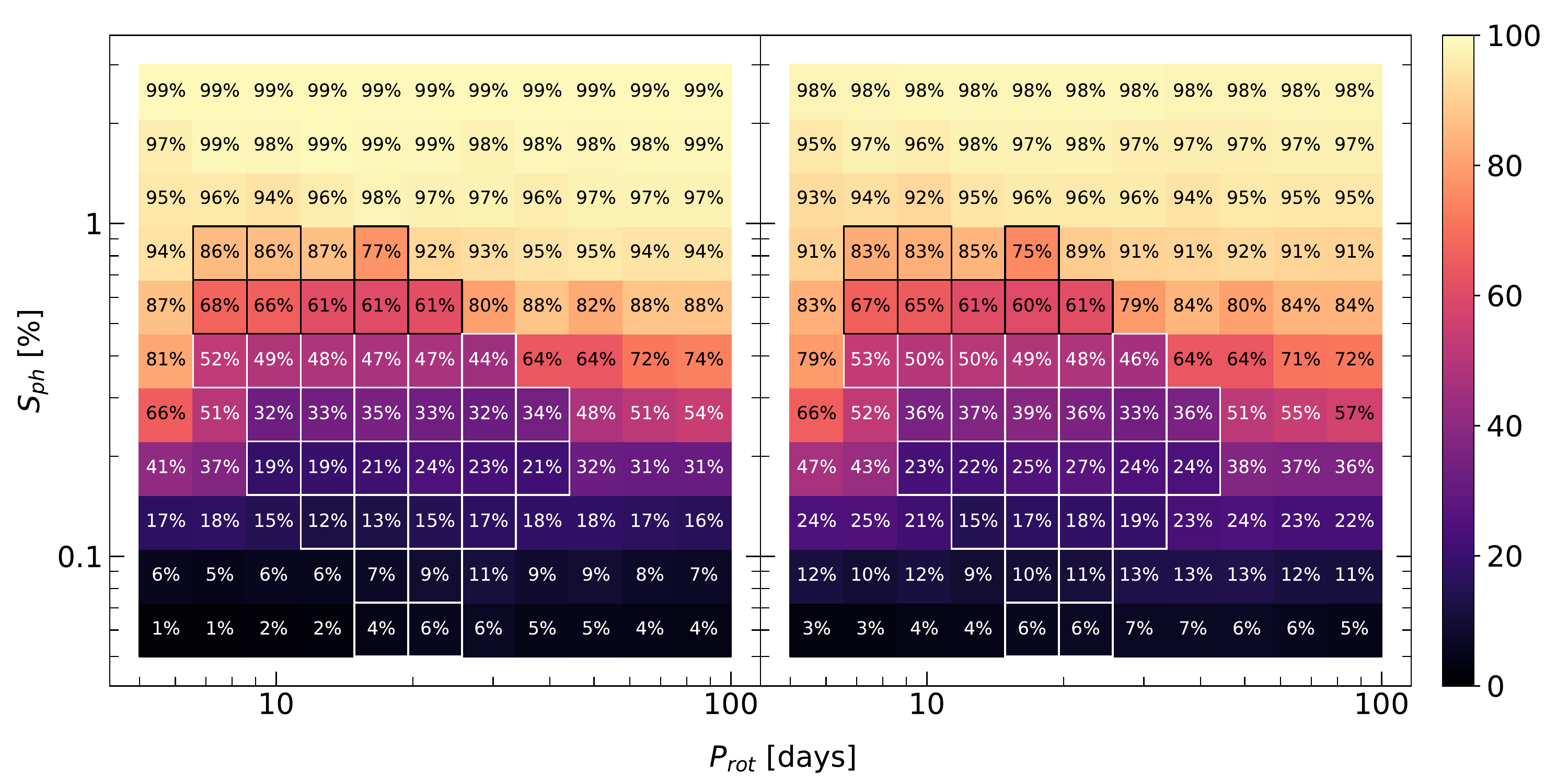}
    \caption{\textbf{Left Panel}: The recovery rate ($\%$) of the period of the signal injected into a cluster member's light curve. \textbf{Right Panel}: Same as left panel, but for field stars rather than cluster members. The gradient from top to bottom demonstrates that the lower a signal's amplitude is, the harder it is to recover. In both panels bins with a black or white (color chosen for optimal contrast) box drawn around them are the ones where over half (i.e.\ $\geq25$) of the injected signals are \textit{Kepler} light curves.}
    \label{fig:iandr-recov}
\end{figure}

\begin{figure}
    \centering
    \includegraphics[scale=0.6]{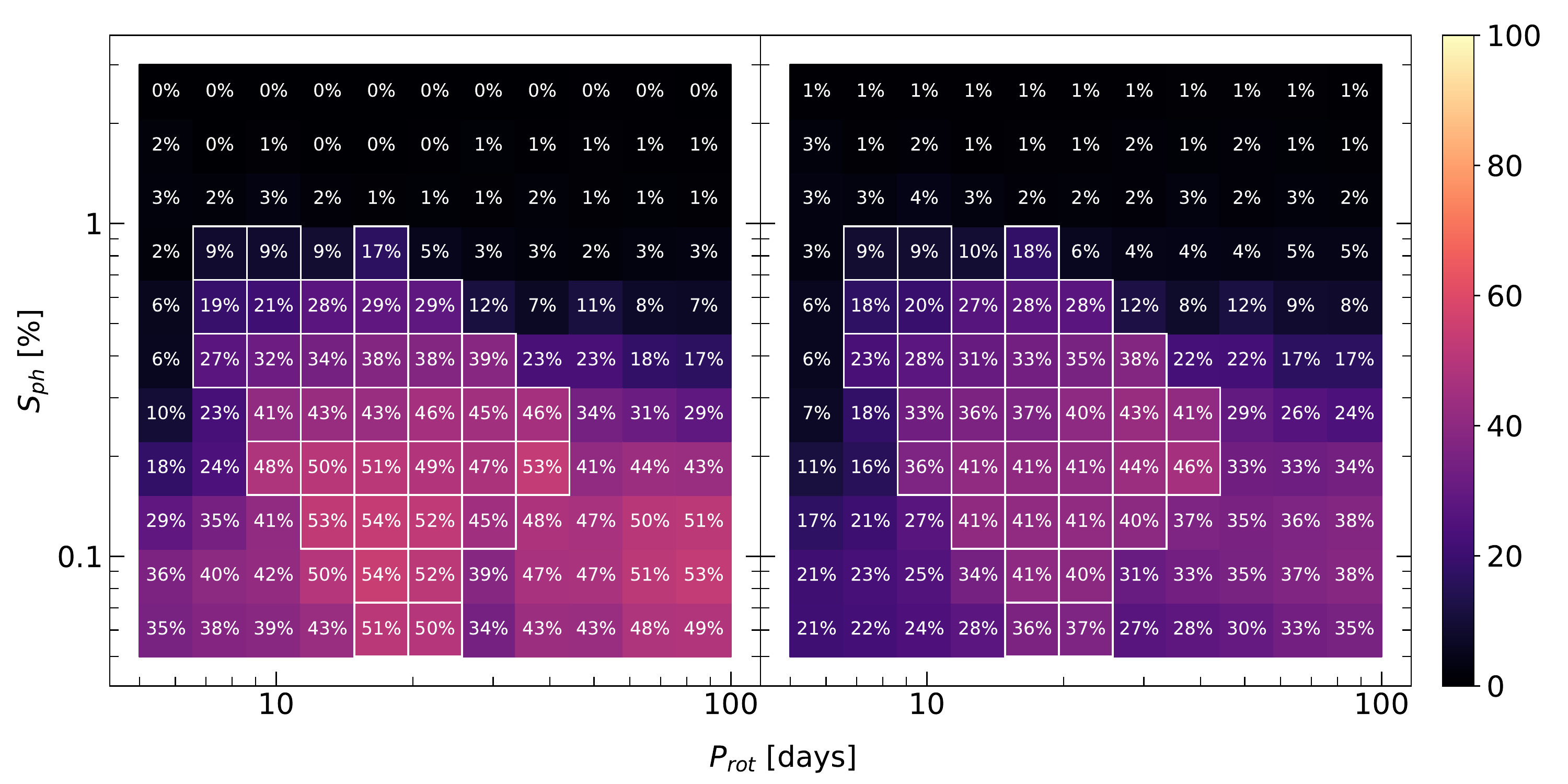}
    \caption{\textbf{Left Panel}: The percentage of the time we find a significant (FAP $< 0.01$) rotation signal but the period does not match what we injected into a cluster member's light curve (to within 10$\%$). \textbf{Right Panel}: Same as the left panel, but for injections into a field star instead of a cluster member. Many of these light curves, especially those of the cluster members, already have periodic signals in them. Thus, in the case of a low amplitude injection we often recover what already existed in the data. Since this does not match what was injected, this is marked as a ``false positive." In both panels bins with a white box drawn around them are the ones where over half (i.e.\ $\geq25$) of the injected signals are \textit{Kepler} light curves.}
    \label{fig:iandr-false}
\end{figure}

We performed injection and recovery tests to determine the detection efficiency and false positive rates for our recovered periods.
The \textit{Kepler} long-cadence data provided a database of real astrophysical signals of rotation for us to test our rotation recovery.
For the injected signals, we used the KEPSEISMIC light curves of K and M dwarf main sequence stars obtained with the \textit{Kepler} Asteroseismic Data Analysis and Calibration Software \citep[KADACS;][]{garcia2011,garcia2014,pires2015}.
The rotation periods for these stars have been derived by \citet{santos2019}.
We applied a few additional cuts of our own: first, we checked for the completeness of the \textit{Kepler} light curve, rejecting any star with fewer than 11 continuous quarters of data.
Second, we applied a cut on the height of the autocorrelation function peak \citep[$H_\mathrm{ACF}$ in ][the average difference between the peak height and the two adjacent local minima]{santos2019}
requiring that $H_\mathrm{ACF}\geq1.0$, a value typical of stable signals.
Finally, we applied a cut on the effective temperature of $T_{\mathrm{eff}} < 5270\;\mathrm{K}$, so that the observed spot pattern evolution in the \textit{Kepler} sample would more closely match the expectation for our targets in M67.
Given the precision of the \textit{Kepler} photometry relative to our data, we made the assumption that these light curves contained noiseless rotation signals.

This gave us a sample of 4599 signals with known rotation periods for injection.
Each signal was characterized by two values: the rotation period ($P_{\mathrm{rot}}$) and the photometric activity index ($S_{\mathrm{ph}}$), a measure of the amplitude of variability.
The value of $S_{\mathrm{ph}}$ was calculated by dividing a light curve into sub-series, each five times the length of the star's rotation period, and then taking the mean of the standard deviations of each of the sub-series \citep{mathur2014}.
One the advantages of $S_\mathrm{ph}$ over other measures of photometric variability is its correlation with proxies of magnetic activity \citep{salabert2016,salabert2017}.
We created logarithmically spaced bins for the injections: $5 \leq P_{\mathrm{rot}} \leq 100\;[\mathrm{days}]$ and $0.05 \leq S_{\mathrm{ph}} \leq 3\; [\% \mathrm{Flux}]$ using 11 bins along each axis.
However, this left some of the outlier bins (see Fig. \ref{fig:kplr_in_bins}) with very few, if any, injections.
To compensate for this deficiency we also generated a set of synthetic light curves.
These synthetic light curves were simple sinusoids:
\begin{equation}
    \mathrm{Flux}(t) = \sqrt{2} S_{\mathrm{ph}} \sin\left(\frac{2\pi}{P_{\mathrm{rot}}} t + \phi\right).
\end{equation}
Where $\phi$ is a uniformly distributed phase, and the factor of $\sqrt{2}$ comes from the fact that $S_{\mathrm{ph}}$ is calculated from a standard deviation.
The synthetic light curves were sampled with the same cadence as the \textit{Kepler} data.
For every bin with less than 50 \textit{Kepler} light curves we generated a sample of up to 50 synthetic ones with $P_{\mathrm{rot}}$ and $S_{\mathrm{ph}}$ values uniformly distributed (in linear space) within the bounds of that bin.
In total we used 3934 synthetic light curves.

Each injection and recovery test involved taking a \textit{Kepler} (or synthetic) light curve and sampling it to match the cadence and length of our CFHT observations.
We then added the signal into one of our CFHT light curves and computed an LS periodogram for the combined data.
If the period of maximum power in the resulting LS periodogram was within $10\%$ of the injected period (see Fig.\ \ref{fig:ls_precision}) and had an estimated FAP of less than $1\%$, then we considered this a successful recovery.
If the period of maximum power was more than $10\%$ different from the injected period and the estimated FAP was less than $1\%$ we considered this a false positive.
All other cases were considered non-detections.
We did not want to assume that the period of maximum power in our periodograms was due to rotation, thus we did not remove any pre-existing signal from the light curves before injection.
To prevent confusion with the signal already present in the CFHT lightcurves we removed any case where the injected period was within $10\%$ of the signal already detected in the lightcurve.
This filtered out no more than $11\%$ of the tests in any given bin, with every bin having at least 45000 tests.
This approach enabled us to incorporate the actual systematics present in our CFHT photometry that may have limited the recovery of periodic signals.

Since we are only interested in the rotation periods of the members of M67, we limited the sample of CFHT light curves to a subset of the candidate cluster members and a matching number of randomly selected field stars.
We selected the light curves for the injection and recovery testing by applying three criteria.
First, cluster members were required to have very high (i.e.\ $=1.0$) membership probability, while field stars must have had very low (i.e.\ $=0.0$) membership probability.
Second, they must have had at least 99 epochs of data available.
Finally, they must not have had an overlapping aperture.
We also required that the selected field stars have similar $r-i$ colors and apparent $i$ magnitudes to our selected cluster members, to mitigate the impact of any systematics that depended on color.
This yielded 740 total light curves, 370 cluster members and 370 field stars, into which we injected each of our 4599 \textit{Kepler} and 3934 synthetic light curves.
Each injection and CFHT light curve pairing was repeated with three or four different phases, depending on how many CFHT light curves fit within the injection light curve.
This was done to capture the shift in phase due to spot pattern evolution over the years of observations.

We compiled the results of these tests into our completeness diagram (Fig.\ \ref{fig:iandr-recov}) as well as our false positives diagram (Fig.\ \ref{fig:iandr-false}).
We have plotted the results from the cluster members and field stars separately.
Injections into the light curves of cluster members served as a direct test of our ability to recover rotation signals in the cluster member data, while injections into the field stars served as a control sample.
The underlying distribution of $P_\mathrm{rot}$ is different for each of these populations, and thus each is expected to impact the completeness diagram in different ways.
Trends that are common to both figures are thus reflective of the pipeline's recovery capabilities in general.
Any differences between the two panels that cannot be attributed to different $P_\mathrm{rot}$ distributions would reflect issues in the pipeline, but we do not see any such differences.

The completeness diagram (Fig.\ \ref{fig:iandr-recov}) shows the major trends we would expect: 1) as the amplitude of the rotation signal decreases our ability to recover the correct period also decreases, and 2) the evolving spot patterns in the \textit{Kepler} light curves reduced our ability to recover the correct period.
Our false positives diagram (Fig.\ \ref{fig:iandr-false}) also shows the major trends that we expected.
In particular we highlight the $10-20\%$ difference in false positive rates between cluster members and field stars for low amplitude injections ($S_\mathrm{ph}\lesssim0.25\%$).
Many of the light curves we injected signals into already had an existing periodic signal and, when injecting low amplitude signals we expected to instead recover the already present signal.
This explains both the high percentage of false positives for low amplitude injections, as well as the difference in false positive rates.
The cluster members were generally expected to show periodic variability due to their spot patterns.
On the other hand, a smaller fraction of field stars were expected to show rotational variability and those that do span a much wider range of timescales (e.g., background evolved stars).

\section{Results and Analysis} \label{sec:res}

\begin{deluxetable}{llcl}
\tablecaption{Catalog of M67 Members \label{tab:results}}
\tablehead{\colhead{Name} & \colhead{Table Header} & \colhead{Units} & \colhead{Description}}
\startdata
\textit{Gaia} Source ID & \texttt{gaiaid} & -- & \textit{Gaia} EDR3 \texttt{source\_id}\\
Right Ascension (RA) & \texttt{RAdeg} & $^\circ$ & --\\
Error on RA & \texttt{e\_RAdeg} & mas & --\\
Declination (Dec) & \texttt{DEdeg} & $^\circ$ & --\\
Error on Dec & \texttt{e\_DEdeg} & mas & --\\
Parallax & \texttt{plx} & mas & --\\
Error on Parallax & \texttt{e\_plx} & mas & --\\
Proper Motion in RA & \texttt{pmRA} & mas/yr & --\\
Error on Proper Motion in RA & \texttt{e\_pmRA} & mas/yr & --\\
Proper Motion in Dec & \texttt{pmDE} & mas/yr & --\\
Proper Motion in Dec & \texttt{e\_pmDE} & mas/yr & --\\
Renormalized Unit Weighted Error & \texttt{ruwe} & -- & --\\
\textit{Gaia} G Magnitude & \texttt{gaiaGmag} & mag & --\\
Error on \textit{Gaia} G Magnitude & \texttt{e\_gaiaGmag} & mag & --\\
\textit{Gaia} BP Magnitude & \texttt{gaiaBPmag} & mag & --\\
Error on \textit{Gaia} BP Magnitude & \texttt{e\_gaiaBPmag} & mag & --\\
\textit{Gaia} RP Magnitude & \texttt{gaiaRPmag} & mag & --\\
Error on \textit{Gaia} RP Magnitude & \texttt{e\_gaiaRPmag} & mag & --\\
PS1 Source ID & \texttt{ps1id} & -- & Pan-STARRS1 DR2 \texttt{ObjID}\\
PS1 $g$ magnitude & \texttt{ps1gmag} & mag & --\\
Error on PS1 $g$ magnitude & \texttt{e\_ps1gmag} & mag & --\\
PS1 $r$ magnitude & \texttt{ps1rmag} & mag & --\\
Error on PS1 $r$ magnitude & \texttt{e\_ps1rmag} & mag & --\\
PS1 $i$ magnitude & \texttt{ps1imag} & mag & --\\
Error on PS1 $i$ magnitude & \texttt{e\_ps1imag} & mag & --\\
PS1 $z$ magnitude & \texttt{ps1zmag} & mag & --\\
Error on PS1 $z$ magnitude & \texttt{e\_ps1zmag} & mag & --\\
Probability of Membership & \texttt{memberprob} & -- & Probability of Membership based on Kinematics (Sec.\ \ref{subsec:membership}) \\
Photometric Single Star & \texttt{single} & -- & Star was determined to be single (Sec.\ \ref{subsubsec:binaries}) \\
Photometric Binary & \texttt{binary} & -- & Star was determined to be a multiple system (Sec.\ \ref{subsubsec:binaries}) \\
Member in Gao's M67 Catalog & \texttt{gaomember} & -- & Star is also listed as a member by \citet{gao2018}\\
Potentially Suspect Member & \texttt{suspect} & -- & Star's membership depended on catalog size (Sec.\ \ref{subsec:membership})\\
Used in Fit & \texttt{converged} & -- & Is used in the polynomial fit after outlier rejection. \\
Effective Temperature & \texttt{teff} & K & Effective temperature derived from ($r-i$) color (Sec.\ \ref{subsec:teffs})\\
Rotation Period & \texttt{prot} & d & Rotation period derived from Lomb-Scargle Periodograms (Sec.\ \ref{sec:rotation}) \\
False Alarm Probability & \texttt{fap} & -- & The estimated false alarm probability of the rotation period \\
\enddata
\tablecomments{A description of the columns in the table of results that is available in a machine-readable format. Astrometric and \textit{Gaia} Photometry are taken from \textit{Gaia} EDR3. PS1 Photometry is taken from PS1 DR2.}
\end{deluxetable}

\begin{figure}
    \centering
    \includegraphics[scale=0.6]{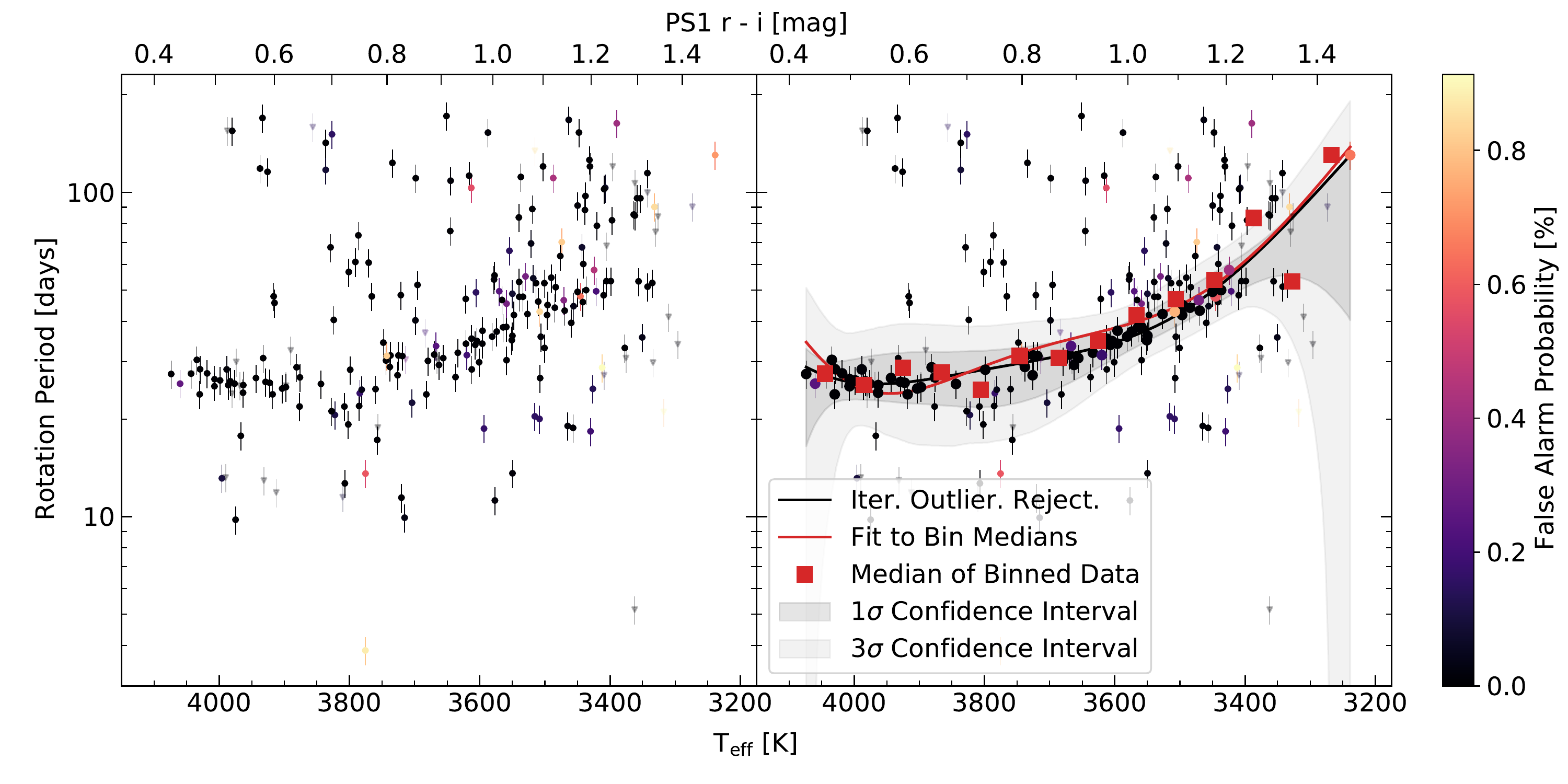}
    \caption{\textbf{Left Panel}: Recovered rotation periods for members of the open cluster M67 plotted versus their color-derived effective temperatures. All points have a low false alarm probability ($<1\%$) for the detected rotation period. \textbf{Right Panel}: We fit a polynomial to the sequence by iterative outlier rejection. The larger points are used in the polynomial fit after outlier rejection. \textbf{The shaded region is the confidence interval of this fit.} Red squares are the median values of the data binned in $T_\mathrm{eff}$, and the red line is a polynomial fit to these values.}
    \label{fig:teff_v_period}
\end{figure}

We present the full rotation catalog in Table \ref{tab:results}. This information includes: \textit{Gaia} EDR3 and PS1 DR2 source identifiers, the \textit{Gaia} EDR3 astrometry and photometry, the PS1 DR2 $griz$ photometry, the recovered rotation periods (if available), their estimated FAP values, the derived $T_\mathrm{eff}$ values, the percentage probability we calculated for M67 membership, and whether or not the star was flagged as a candidate binary.
Appendix \ref{extra_figures} contains plotted light curves and periodograms for each star in Table \ref{tab:results} with a reported rotation period.

We have plotted the measured rotation periods versus effective temperature for the 294 candidate single members of M67 with significant rotation detections in Fig.\ \ref{fig:teff_v_period}.
For the analysis, we applied two extra cuts on the rotation periods, requiring that the stars have not been flagged as having an overlapping aperture (Sec.\ \ref{subsec:phot}) or as a field-of-view dependent member (Sec.\ \ref{subsec:membership}).
Despite some scatter in the periods, they are concentrated about a locus in $T_\mathrm{eff}$-$P_\mathrm{rot}$ space.
In an effort to describe this sequence, we performed a polynomial fit to the data using iterative outlier rejection, where at each step outliers were defined as the data greater than three median absolute deviations away from the median of the residuals.
We did this for both $T_\mathrm{eff}$ vs $P_\mathrm{rot}$ and PS1 $(r-i)$ vs $P_\mathrm{rot}$, finding that both approaches converged to the same solution: a subset of 64 stars, for which the least-squares best fits are:
\begin{equation}
\label{eq:thefit}
    P_\mathrm{rot}(T_\mathrm{eff,4K}) = 9.66\times10^{-10} \cdot T_\mathrm{eff,4K}^4 + 8.25\times10^{-7} \cdot T_\mathrm{eff,4K}^3 +  2.69\times10^{-4} \cdot T_\mathrm{eff,4K}^2 + 0.016 \cdot T_\mathrm{eff,4K} + 25.9,
\end{equation}
or
\begin{equation}
\label{eq:colorfit}
    P_\mathrm{rot}(r-i) = 292 \cdot (r-i)^4 - 895 \cdot (r-i)^3 +  1054 \cdot 
    (r-i)^2 - 543 \cdot (r-i) + 127.9,
\end{equation}
where $T_\mathrm{eff,4K}= T_\mathrm{eff} - 4000\;\mathrm{K}$.
We used bootstrapping (N=10000) to calculate confidence intervals about our fit, fitting a polynomial to 64 stars sampled with replacement from the 253 stars that passed all quality cuts.
As a point of comparison, we have also taken the approach of binning the 253 stars in $T_\mathrm{eff}$, computing a median $P_\mathrm{rot}$ for each bin, and fitting a polynomial to these medians.
The medians are plotted as red squares in Fig.\ \ref{fig:teff_v_period}, and their fit is plotted as a red line, which we have found is in agreement with the iterative outlier approach.
We favor the results of the iterative outlier rejection due to its exclusion of points we believe are aliases from the fit to the $T_\mathrm{eff}$ vs $P_\mathrm{rot}$ sequence (see related discussion in Sec.\ \ref{subsec:uncer}).

\subsection{Lomb-Scargle Failure Modes} \label{subsec:failmodes}

\begin{deluxetable}{llcl}
\tablecaption{Detected Lomb-Scargle Failure Modes \label{tab:failmodes}}
\tablehead{\colhead{Name} & \colhead{Table Header} & \colhead{Units} & \colhead{Description}}
\startdata
\textit{Gaia} Source ID & \texttt{gaiaid} & -- & \textit{Gaia} EDR3 \texttt{source\_id} \\
False Alarm Probability & \texttt{fap} & -- & The estimated false alarm probability of the rotation period \\
Rotation Period ($P_\mathrm{rot}$) & \texttt{prot} & d & The measured rotation period of the star \\
Half Period Alias & \texttt{m=2} & d & $m=2$, and $n=0$ \\
Third Period Alias & \texttt{m=3} & d & $m=3$, and $n=0$ \\
First Month Failure Mode & \texttt{monthn=-2} & d & $m=1$, $n=-2$, and $\delta P\approx29.5$ Days \\
Second Month Failure Mode & \texttt{monthn=-1} & d & $m=1$, $n=-1$, and $\delta P\approx29.5$ Days \\
Third Month Failure Mode & \texttt{monthn=+1} & d & $m=1$, $n=1$, and $\delta P\approx29.5$ Days \\
Fourth Month Failure Mode & \texttt{monthn=+2} & d & $m=1$, $n=2$, and $\delta P\approx29.5$ Days \\
First Year Failure Mode & \texttt{yearn=-2} & d & $m=1$, $n=-2$, and $\delta P\approx380.8$ Days \\
Second Year Failure Mode & \texttt{yearn=-1} & d & $m=1$, $n=-1$, and $\delta P\approx380.8$ Days \\
Third Year Failure Mode & \texttt{yearn=+1} & d & $m=1$, $n=1$, and $\delta P\approx380.8$ Days \\
Fourth Year Failure Mode & \texttt{yearn=+2} & d & $m=1$, $n=2$, and $\delta P\approx380.8$ Days \\
\enddata
\tablecomments{A description of the columns that are in the machine readable table. All alias and failure mode values are the output of Eq.\ \ref{eq:lsfail} with $P_\mathrm{true}=P_\mathrm{rot}$ and are only included in the table if a peak was detected at that period in the periodogram.}
\end{deluxetable}

\begin{figure}
    \centering
    \includegraphics[scale=0.6]{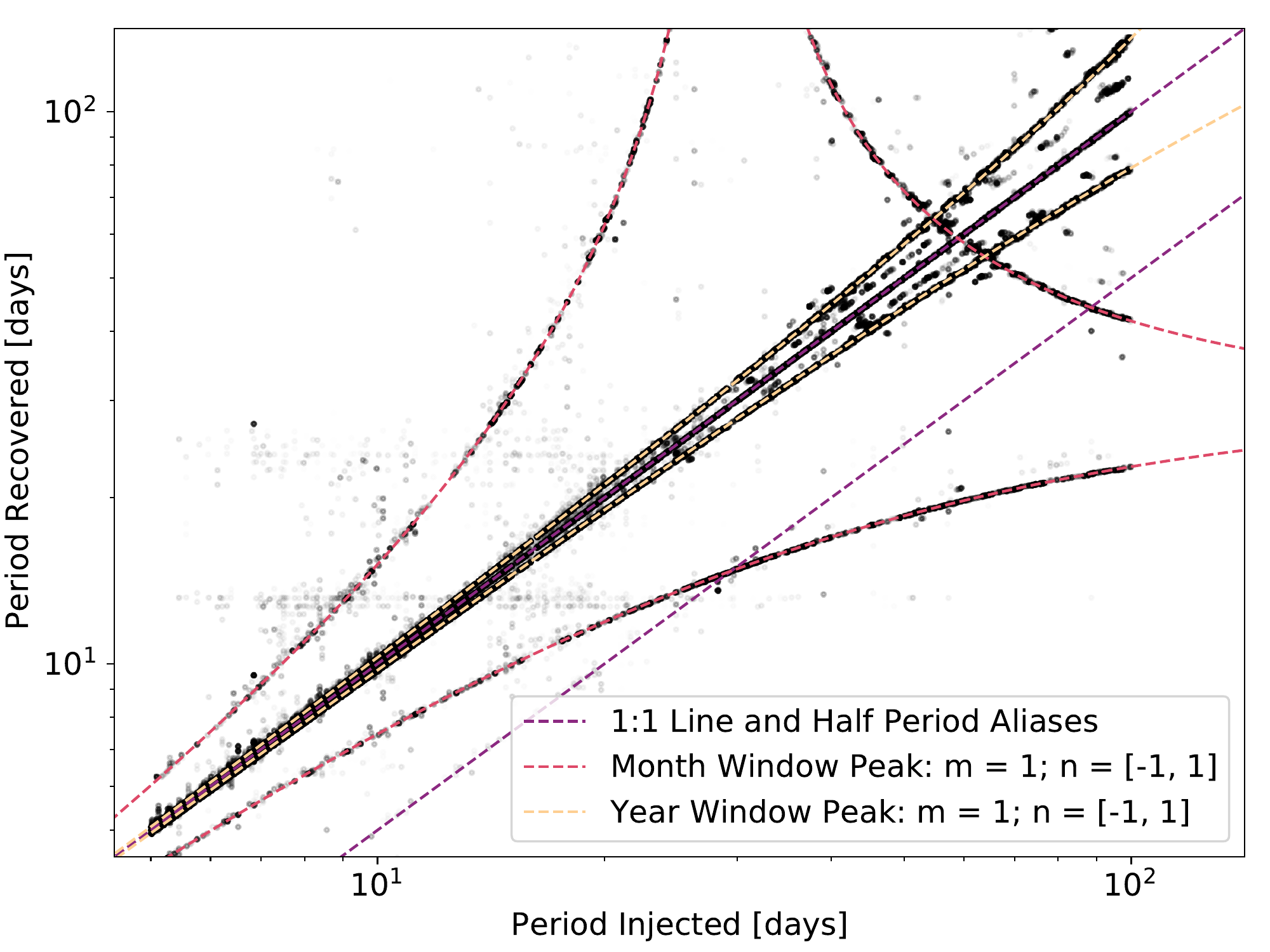}
    \caption{A scatter plot of the injected period vs the recovered period for one CFHT light curve. Each point represents the result of one \textit{Kepler} (or synthetic) light curve being added to this cluster member's light curve. Low amplitude injections are dominated by the existing signal in the data, resulting in the horizontal features on this diagram. Dashed lines represent the 1-to-1 line for successful recoveries and the most prominent window function effects: half period aliasing and the Month and Year window peaks (see Sec.\  \ref{subsec:failmodes} for detailed discussion). Additional peaks in the window function align with other trends in this figure but are not plotted to reduce figure crowding (e.g., $m=1$, $n=\pm1$ and $\delta P\sim180$ would fall between the plotted year-based effect and the 1:1 line).}
    \label{fig:iandr_scatter}
\end{figure}

In addition to the $10\%$ uncertainties determined from our injection and recovery tests (Sec.\ \ref{subsec:iandr}) there are systematic uncertainties contributing to the scatter in our results (Fig.\ \ref{fig:teff_v_period}).
These are the failure modes of the LS periodogram, originating from the irregular sampling in time.
For the purposes of this discussion we will be using the term ``window function" in a slightly different manner than in more classical time series analysis discussions.
Instead of describing a traditional window function, such as the Hann window, we take an approach similar to that of \citet{vanderplas2018} where the window function describes how the light curve was sampled in time.
This window function has predictable effects on the LS periodograms computed from the data, which can all be combined into one equation \citep[Eq.\ 47 in][]{vanderplas2018}:
\begin{equation}
    \label{eq:lsfail}
    P_\mathrm{obs} = \left| \frac{m}{P_\mathrm{true}} + \frac{n}{\delta P} \right|^{-1},
\end{equation}
where $P_\mathrm{obs}$ is the observed peak in the periodogram, $P_\mathrm{true}$ is the true period of the underlying signal, and $m$ and $n$ are integers.
$m=1$ and $n=0$ yields the true period, and $m=2$ and $n=0$ represents the classic case of half-period aliasing, however they can both take any integer value, positive or negative.
The final term, $\delta P$, is the period of a peak in the window function's periodogram; there are typically more than one, and for our CFHT observations there were two dominant ones.
They were the ``month window peak" ($\delta P \approx 29.5$ Days) arising from only observing during the bright lunar phases, and the ``year window peak" ($\delta P \approx 380.8$ Days) arising from only observing when the cluster is up.
The exact values of each depends on the precise sampling in time (i.e., on the completeness of the light curve).
The effects of these window peaks can be easily seen in a scatter plot of $P_\mathrm{true}$ vs $P_\mathrm{obs}$, which we have plotted using the results of a complete set of injections into one of our CFHT light curves (Fig.\ \ref{fig:iandr_scatter}).
There is no way to determine if the period of maximum power in a periodogram corresponds to $P_\mathrm{true}$ or one of its failure modes, $P_\mathrm{obs}$, with absolute certainty.
Moreover, because $m$ and $n$ are integers there is no continuum of window effects, meaning a standard deviation computed across the peaks in a periodogram is a poor description of the uncertainty.
\citet{vanderplas2018} provide a prescription for how one can use detected failure modes to improve the accuracy of interpreting periodgrams.
We did not use their prescription, instead we found that many of the stars identified as likely to be failure modes as opposed to true rotation periods by their method are rejected in our iterative-outlier rejection and thus already excluded from our analysis.
We have included a table of all the detected potential failure modes associated with each of our reported $P_\mathrm{rot}$ values in Table \ref{tab:failmodes}.
Readers who are interested in further trimming to create their own subset of the M67 rotation periods reported here may use these values in a prescription like that of Sec.\ 7.2 of \citet{vanderplas2018}.

\subsection{Deviations from the Sequence}
\label{subsec:uncer}

\begin{figure}
    \centering
    \includegraphics[scale=0.6]{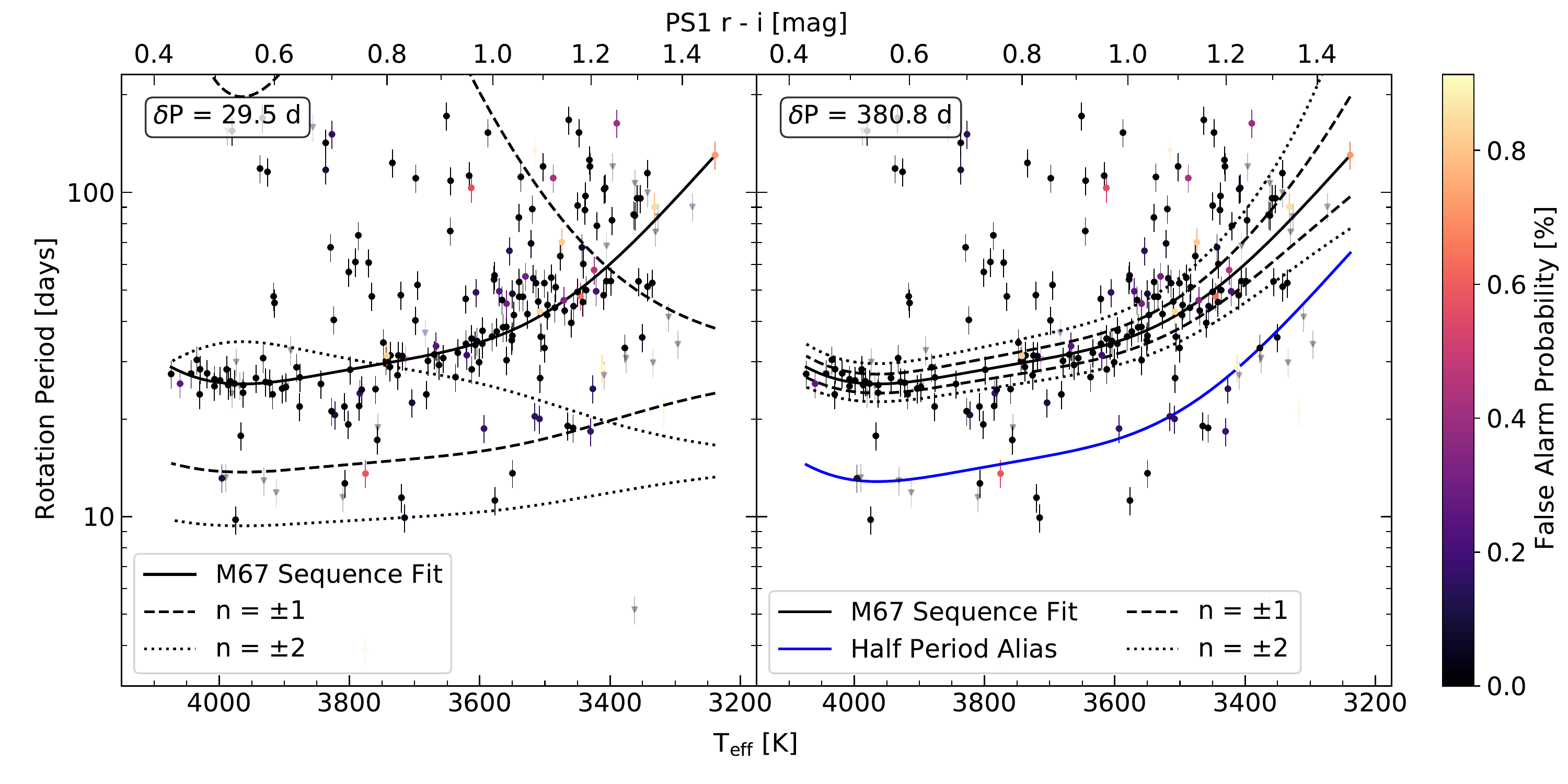}
    \caption{\textbf{Both Panels}: The rotation period vs effective temperature of M67 members with our fit to the M67 sequence (solid black). Dashed lines are the result of assuming our fit to the sequence is the true period in Eq.\ \ref{eq:lsfail} along with $m=1$ and $n=\pm1$. Dotted lines are the same, but with $n=\pm2$. \textbf{Left Panel}: Using 29.5 Days for $\delta P$ Eq.\ \ref{eq:lsfail}. \textbf{Right Panel}: Using 380.8 Days for $\delta P$ Eq.\ \ref{eq:lsfail}. Additionally the half-period alias of Eq.\ \ref{eq:thefit} is in solid blue.}
    \label{fig:teff_v_windowfunc}
\end{figure}

There are a number of mechanisms that can result in an incorrectly measured value of $P_\mathrm{rot}$, both observational and astrophysical.
Astrophysically, spot pattern evolution can spread the power from a rotation signal into multiple peaks in the periodogram, as well as shift the central location of these peaks.
By using \textit{Kepler} light curves as a part of our injection and recovery tests (Sec.\ \ref{subsec:iandr}) we are able to quantify the effect this has on our recovery.
Comparing the bins with majority \textit{Kepler} light curves to their neighbors with majority synthetic light curves in Fig.\ \ref{fig:iandr-recov} indicates that spot pattern evolution among the \textit{Kepler} light curves lead to a $\sim$15-20$\%$ drop in recovery.
The same comparison using Fig.\ \ref{fig:iandr-false} shows an equivalent uptick in false positives, highlighting the impact of spreading the power across multiple peaks in the periodogram.
Additionally, \citet{basri2018} have shown that stars with lower $T_\mathrm{eff}$ and longer $P_\mathrm{rot}$ tend to favor a ``double dip'' spot pattern that lends itself to half-period aliasing.
However, the results of our injection and recovery tests suggest this is a relatively minor effect for our data set (see the half period alias line in Fig.\ \ref{fig:iandr_scatter}).
Finally, close binary systems will have rotation periods that appear as outliers in the data.
Such systems are affected both by the confusion of brightness modulations in both stars as well as tidal forces changing their rotational evolution relative to single stars.
We have mitigated the contamination from binary stars through our CMD cuts (Sec.\ \ref{subsubsec:binaries}).
Observationally, we were also limited by our irregular sampling in time, such effects are described by the Lomb-Scargle failure modes (see Sec.\ \ref{subsec:failmodes}).
This means that even a perfectly stable sinusoid can be recovered incorrectly as the signal-to-noise ratio on the data decreases.
With these effects in mind we believe that the use of iterative outlier rejection for our reported fit to the $T_\mathrm{eff}$ vs $P_\mathrm{rot}$ sequence was justified.

To demonstrate this, we substituted Eq.\ \ref{eq:thefit} for $P_\mathrm{true}$ in Eq.\ \ref{eq:lsfail}, and plotted the resulting sequences alongside our original results in Fig.\ \ref{fig:teff_v_windowfunc}.
Given the difference in the effect of each window peak we plotted the month and year effects separately.
For the collection of stars rotating faster than our fit to the $T_\mathrm{eff}$ vs $P_\mathrm{rot}$ sequence, they align well with two possible cases: 1) they fall along the half period alias of our fit, or 2) they fall along a sequence that is associated with the month window peak (left panel of Fig.\ \ref{fig:teff_v_windowfunc}).
The sequences associated with the year window peak show that these failure modes contribute to the scatter about our fit to the sequence.
This, combined with the precision on our $P_\mathrm{rot}$ values derived from the injection and recovery tests (Fig.\ \ref{fig:ls_precision}), are what prevent us from measuring a $T_\mathrm{eff}$ vs $P_\mathrm{rot}$ sequence that is as sharply defined as the slow rotator sequences observed in younger clusters \citep[e.g., Praespe and NGC 6811;][]{douglas2017,douglas2019,curtis2019}.

Finally, there remain a number of stars with relatively long rotation periods ($P_\mathrm{rot}\gtrsim40\;\mathrm{d}$) and high temperatures ($T_\mathrm{eff} \gtrsim 3600\;\mathrm{K}$) that are inconsistent with our fitted sequence and its expected failure modes.
To gain some insight into the origin of these inconsistent stars we binned our $P_\mathrm{rot}$ values and computed the fraction of stars in a period bin that were inconsistent with the fitted sequence.
Then we estimated the $S_\mathrm{ph}$ values for these stars, allowing us to compare the computed fractions to the false positive rates of Fig.\ \ref{fig:iandr-false}.
The stars inconsistent with the sequence have a mean $S_\mathrm{ph}$ of $0.3\%$ with a standard deviation of $0.2\%$, compared to a mean $S_\mathrm{ph}$ of $0.4\%$ with a standard deviation of $0.1\%$ for the stars of equivalent temperature consistent with the sequence.
Given these $S_\mathrm{ph}$ values, the inconsistency fractions of $20-50\%$ align well with the false positive rates.
Moreover, the inconsistent stars all have detected failure modes that are consistent with the fitted sequence itself, whereas those on the sequence primarily have detected failure modes consistent with a half-period alias.
A multi-term model would aid in clarifying the true rotation period of these stars, however the sparse sampling in our data made those fits poorly conditioned.
Therefore we are satisfied with their exclusion from our fit to the sequence.
We postulate that these stars were affected by either spot pattern evolution or the signal-to-noise ratio of the data, both of which contribute to how likely a failure mode is to be recovered instead of the true rotation period.
Confirmation would require follow-up observations at much higher cadence and with more regular sampling in time.

\section{Discussion} \label{sec:discuss}

\begin{figure}
    \centering
    \includegraphics[scale=0.6]{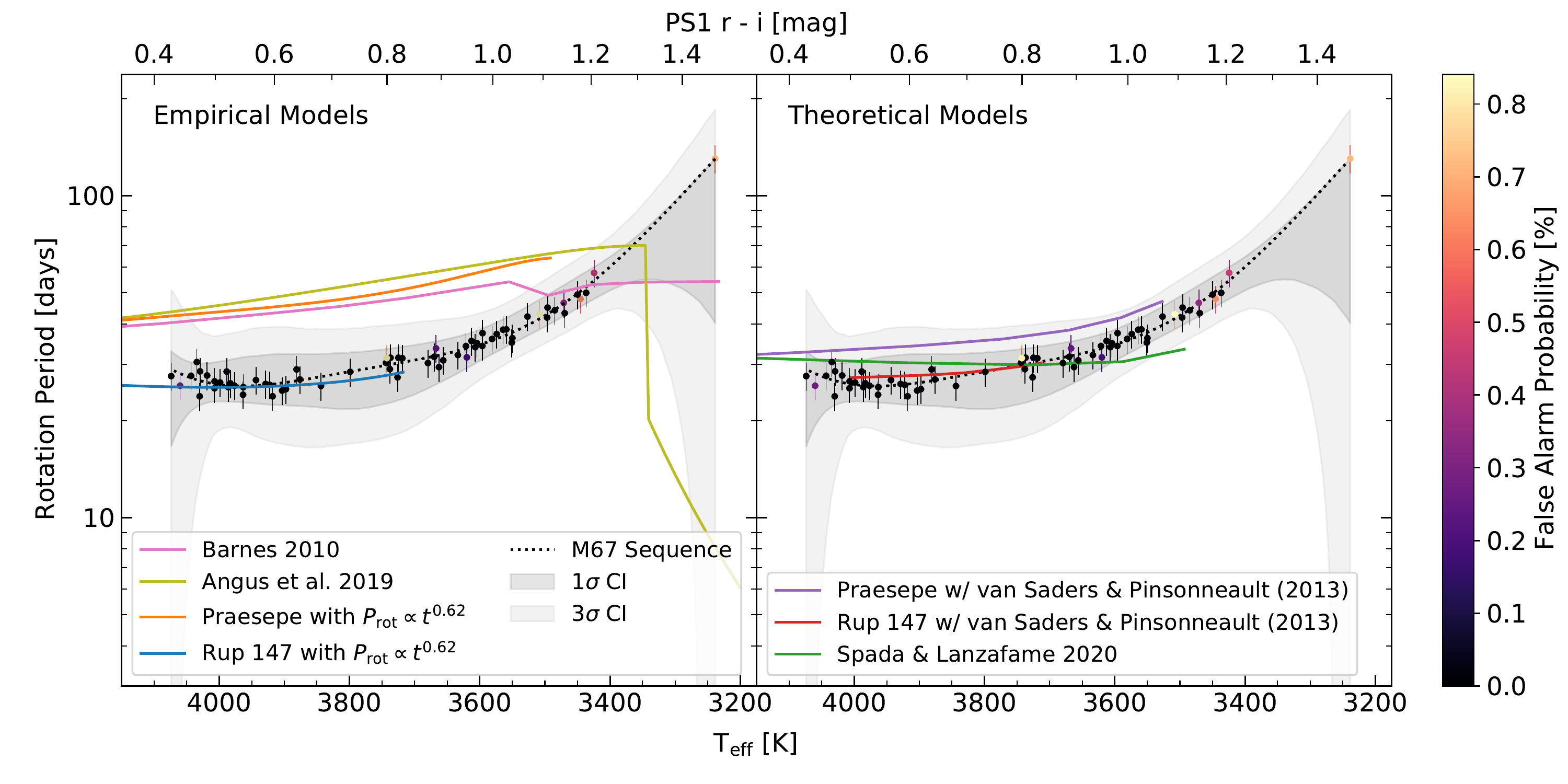}
    \caption{\textbf{Left Panel}: The selected subset (from iterative outlier rejection, Sec.\ \ref{sec:res}) of M67 rotation periods versus their effective temperatures. The shaded region in gray is the confidence interval of the fit, plotted as a dotted black line. Predictions for the sequence from the empirical models are plotted as colored lines. \textbf{Right Panel}: The same as the left panel, but with predictions from the theoretical models.}
    \label{fig:model_compare}
\end{figure}

\begin{figure}
    \centering
    \includegraphics[scale=0.55]{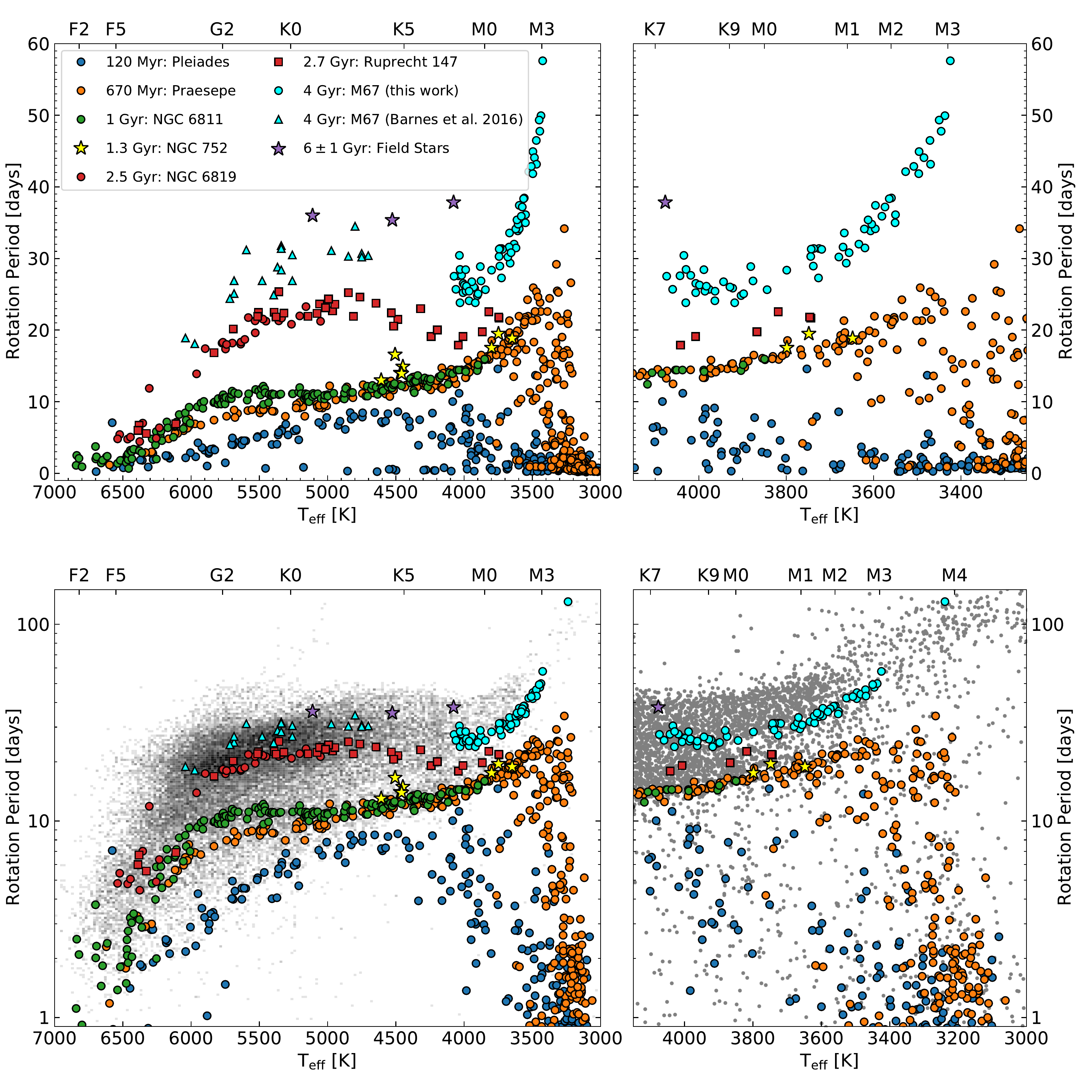}
    \caption{\textbf{Top Left Panel}: A replication of the first panel of Fig. 7 of \citet{curtis2020}, but with our results and the results of \citet{barnes2016} added. \textbf{Top Right Panel}: A Subset of the left panel, zoomed in on the $T_\mathrm{eff}$ range of the results presented in this paper.\textbf{Bottom Left Panel}: Same as top left panel, but now plotted over a 2D histogram of field star rotation periods. \textbf{Bottom Right Panel}: Same as top right panel, but now plotted over field star rotation periods.
    In all cases, the colored points are a collection of open clusters which have been used as gyrochronological benchmarks. Benchmarks include the Pleiades \citep[120 Myr;][]{rebull2016}, Praesepe \citep[670 Myr;][]{douglas2017,douglas2019}, NGC 6811 \citep[1 Gyr;][]{curtis2019}, NGC 752 \citep[1.4 Gyr;][]{agueros2018}, NGC 6819/Ruprecht 147 \citep[2.5 Gyr projected forward by Curtis et al./2.7 Gyr;][]{meibom2015,curtis2020}, M67 \citep[4 Gyr;][and this work]{barnes2016}, and three field stars: $\alpha$ Cen B and 61 Cyg A and B \citep[Table 3 of][and references therein]{curtis2020}. Field stars are taken from a collection of literature sources: \textit{Kepler} \citep{santos2019}, the PS1 Medium Deep Survey \citep{kado-fong2016}, MEarth \citep{newton2016,newton2018}, CARMENES \citep{diezalonso2019}, Evryscope \citep{howard2020}, and the K2SDSS sample \citep{popinchalk2021}.}
    \label{fig:compendium}
\end{figure}

We first compared our observations (the 64 stars our iterative outlier rejection converged to, Sec.\ \ref{sec:res}) against the predictions of two classes of gyrochronological models: empirical and theoretical (Fig.\ \ref{fig:model_compare}).
Empirical models are agnostic to the physics of magnetic breaking and anguluar momentum transport, fitting a relation to a set of periods and ages often as a function of color.
Generally they follow a Skumanich-like relation of $P_\mathrm{rot} \propto t^n$, using cluster data and the Sun as anchor points to tune the value of the exponent.
On the other hand, theoretical models make assumptions regarding the underlying physics and spin-down that manifests from their description.
Whether they are empirical or theoretical in nature, all models are calibrated against objects of known age and rotation period, relatively few of which are young M dwarfs, meaning that their predictions for the age of M67 are an extrapolation.

There are four empirical relations included in our comparison.
First is the \citet{barnes2010} model, where $dP_\mathrm{rot}/dt$ is parameterized in terms of the Rossby number ($Ro$) and two dimensionless constants calibrated on the Sun and young open cluster observations.
Second, the \citet{angus2019} empirical relation, which is a broken power law with mass fit to the sequence of Praesepe and a spin-down law tuned to replicate the Sun.
Finally, we evolved the sequences of Praesepe \citep[670 Myr;][]{douglas2017,douglas2019} and Ruprecht 147 \citep[2.7 Gyr;][]{curtis2020} forward in time through the use of a simple Skumanich-like spin down: $P_\mathrm{rot}(4\;\mathrm{Gyr}) = P_\mathrm{rot,0} (4\;\mathrm{Gyr}/t_0)^{0.62}$.
For the hotter stars in our sample all of these empirical relations, with one exception, predict that the stars of M67 should be rotating $\sim10$-20 days slower than observed.
The exception is the Skumanich-like spin down relation launched from the stars of Ruprecht 147, which provides an excellent match to our data for the earlier M dwarfs ($T_\mathrm{eff} \gtrsim 3700$ K), for the later M dwarfs there were no data available in Ruprecht 147.

The first of the two theoretical models we considered is from \citet{vanSaders2013} and described therein.
Similar to our Skumanich-like empirical relation, we launched this model from two starting places: Praesepe and Ruprecht 147.
This model is calibrated to match the spin down of solar mass stars and the Sun, and then applied to our low-mass M67 members.
Braking laws of this form perform well on the Sun but fail to capture the rotational evolution during the first few hundreds of Myrs in young clusters \citep[see][]{douglas2017,breimann2021,roquette2021}; it is unclear to what degree the issue is with the braking law itself (saturation, core envelope decoupling, etc.) or with the assumed distribution of initial rotation periods \citep{roquette2021}.
We manage these early-time uncertainties by starting our models as solid body rotators at Praesepe age and rotation rate, and evolving them forward to M67.
The model evolved forward from Praesepe is a better match than the empirical relations, as it is within the $3\sigma$ confidence interval, but still predicts that the stars of M67 should be rotating slower than we observed.
The model evolved forward from Ruprecht 147 provides an excellent match, being very closely aligned with the empirical relation launched from the same starting point, although this is again limited by a lack of later M dwarfs in the Ruprecht 147 data.

The second theoretical model we included is the model of \citet{spada2020}.
This iteration on the model includes some minor adjustments from that of \citet{lanzafame2015}, a core-envelope decoupling model.
Their model incorporates a mass-dependent wind-braking law that follows the classical rotation rate dependence of \citet{kawaler1988}, $\frac{dJ}{dt} \propto \omega^3$.
It also uses a two-zone approach to the interior, treating the core and envelope as two separate rotationally solid bodies that are allowed to exchange angular momentum.
As a result, this model explains the apparent stalling of spin-down as a epoch during which significant angular momentum transport occurs from the core to the envelope, balancing out the angular momentum the envelope loses to wind-braking.
The prediction of this model agrees with the observations quite well down to a $T_\mathrm{eff}$ of around 3600 K.
At this point, the model's prediction is too fast compared to our observations, though still consistent at the $3\sigma$ level.

In total, we have found three models that provide an excellent (i.e., within $1\sigma$) match to our observations of M67.
Of these models, two are solid body spin-down (one theoretical, one empirical) applied to the stars of Ruprecht 147, which we interpret as a sign that the late stages of low-mass stellar spin-down are dominated by solid body rotation.
The other model is that of \citet{spada2020}, which is both the only core-envelope decoupling model tested, as well as the only model launched from the birth of the star.
The excellent agreement between it and our observations makes a compelling case for the core-envelope decoupling theory.

\subsection{The Case for Core-Envelope Decoupling} \label{subsec:thecase4ced}
The evidence for core-envelope decoupling goes deeper than the agreement between the model of \citet{spada2020} and our observations.
In the core-envelope decoupling framework, after the epoch of significant angular momentum transport occurs, the expectation is that the core and envelope of the star have equalized in angular velocity.
At this point, the star spins down as a solid body.
If the stars of Ruprecht 147 have resumed their spin-down \citep{curtis2020}, core-envelope decoupling would predict that they are now spinning down as solid bodies.
The precise agreement between our solid body and empirical models launched from Ruprecht 147 and our observations implies that this is the case, at least through the age of M67.

Another important test of core-envelope decoupling lies in the behavior of spin-down for stars that are nearly or fully convective.
The diminishing size of the core limits the amount of angular momentum it can store relative to the envelope, reducing the length of time the star's spin-down would stall.
Furthermore, stars with no radiative core should not stall their spin-down at all.
\citet{curtis2020} provide an empirical relation for the age at which stars resume spinning down. They find:
\begin{equation} 
\label{eq:resume}
    t_\mathrm{R} = 0.202\;\mathrm{Gyr} \times \left(\frac{T_\mathrm{eff}}{5770\;\mathrm{K}}\right)^{-5.11},
\end{equation}
based on a simplified model where spin-down comes to a full stop and then suddenly resumes after a mass-dependent length of time.
If we extrapolate this relation to later spectral types (i.e., beyond M0), we find that by M1.5 the age at which spin down resumes is approaching that of M67 (3.6 Gyr vs 4 Gyr), and by M2 this age has potentially exceeded that of M67 (4.6 Gyr).
However, the stars in this range of temperatures (M1-M3) are rotating $\sim$10-30 days slower than their younger counterparts (top two panels of Fig.\ \ref{fig:compendium}), implying they have been spinning-down for at least part of the intervening $\sim3$ Gyrs.
This is suggestive of a need for a $t_\mathrm{R}$ relation that has a turnover as it approaches the fully convective boundary, as expected in the core-envelope decoupling framework.
Observations of younger M dwarfs of these spectral types (e.g., those in Ruprecht 147 or NGC 752) will be a critical test for determining when these stars resumed their spin-down.

\subsection{M67 and the Field} \label{subsec:M67inthefield}
We have also compared our observations to an ensemble of field star rotation periods collected from a variety of sources.
The largest contributor to this collection is the \textit{Kepler} sample, with temperatures and rotation periods from \citet{santos2019}.
The rest are predominantly M dwarfs with rotation periods from: the PS1 Medium Deep Survey \citep{kado-fong2016}, MEarth \citep{newton2016,newton2018}, CARMENES \citep{diezalonso2019}, Evryscope \citep{howard2020}, and the K2SDSS sample \citep{popinchalk2021}.
\citet{popinchalk2021} provided the \textit{Gaia} DR2 identifiers for the targets from all of these surveys, which we used to obtain temperatures from v8 of the \textit{TESS} Input Catalog (TIC) \citep{stassun2019}.
We then plotted these field stars with the open cluster data (bottom two panels of Fig.\ \ref{fig:compendium}), ignoring any stars which did not have a temperature in TIC.

Field M dwarfs follow a bimodality in their rotation periods \citep{kado-fong2016,newton2016,howard2020}.
Using kinematic ages, \citet{newton2016} speculated that the transition between the fast and slow populations must be quick, and must occur between the ages of 2 and 5 Gyr.
The stars of M67 fall along the lower envelope of the slow rotator population, suggesting that they represent convergence onto a slow rotator sequence for M dwarfs.
The age of M67 (4 Gyr) is consistent with the bounds for the transition.
Higher cadence observations are needed to confirm whether or not there are still rapid rotators in M67, a lack of which would make M67 a fully converged slow rotator sequence.
Since accurate gyrochronology depends on stars converging to a slow rotator sequence, the age of M67 serves as a lower bound for accurate gyrochronological ages of M dwarfs.

Another interesting feature seen in the distribution of field star rotation periods is the intermediate period gap.
This is a bimodal distribution of stars with $T_\mathrm{eff}$ values less than 5000 K and intermediate rotation periods \citep[15-25 Days,][]{mcquillan2013}.
A number of explanations have been put forth to explain this gap, including a lull in star-formation \citep{davenport2017}, a transition to faculae-dominated photospheres \citep{reinhold2019}, or an epoch of accelerated spin-down during the recoupling of the core and envelope \citep{mcquillan2013,gordon2021}.
Open cluster data shows that any explanation relying on the gap stars having a common age is incorrect \citep{curtis2020}.
Instead the mechanism that causes this gap must occur at different times for stars of different masses.
The stars of M67 appear along the upper envelope of the intermediate period gap, suggesting an upper bound of 4 Gyr for the age by which this mechanism has occurred.
Furthermore, if the gap is indeed caused by accelerated spin-down during core-envelope recoupling, then the stars along the upper envelope of the intermediate period gap should be composed of stars that are spinning down as solid bodies, in line with our observations.

\subsection{Evidence of a Unique Spin-Down History} \label{subsec:cautionadvised}
While this description is compelling, some caution is important.
\citet{somers2016} identified M67 as an outlier among open clusters in terms of its lithium abundance.
Having demonstrated that Li depletion is a strong test of core-envelope recoupling they concluded that the most likely scenario explaining M67's Li abundances is an ``intrinsically different mixing history" driven by a surplus of rapid rotators in the cluster's early years.
Observations of young clusters and associations show that massive stars in large clusters can drive photoevaporation of the disks of nearby lower mass stars, shortening disk lifetimes and resulting in a larger population of rapid rotators \citep{roquette2021}.
Such a surplus of rapid rotators would shift the mean sequence of M67 to faster rotation periods compared to stars of equivalent ages until the initial conditions are forgotten.
However, this will not affect the braking laws describing their spin-down.
We can control for M67's unique initial rotation periods by modeling a variety of cases for the initial conditions, as well as observing other clusters of similar ages.

\section{Conclusions} \label{sec:conclusion}

In this paper we have:
\begin{itemize}
    \item Generated a new catalog of 1807 M67 members based on \textit{Gaia} EDR3 parallaxes and proper motions, identified potential unresolved binaries by their location on the cluster's CMD, and calculated the color-based effective temperatures for the late K and early M dwarf single members of M67.
    \item Reported the rotation periods for 294 of these M67 members, providing a sample of 4 Gyr old late K and early M dwarfs for calibrating gyrochronological models and a polynomial fit to the sequence they form in $T_\mathrm{eff}$ vs $P_\mathrm{rot}$ for use as a gyrochrone:
    \begin{equation}
    P_\mathrm{rot}(T_\mathrm{eff,4K}) = 9.66\times10^{-10} \cdot T_\mathrm{eff,4K}^4 + 8.25\times10^{-7} \cdot T_\mathrm{eff,4K}^3 +  2.69\times10^{-4} \cdot T_\mathrm{eff,4K}^2 + 0.016 \cdot T_\mathrm{eff,4K} + 25.9,
\end{equation}
or
\begin{equation}
    P_\mathrm{rot}(r-i) = 292 \cdot (r-i)^4 - 895 \cdot (r-i)^3 +  1054 \cdot 
    (r-i)^2 - 543 \cdot (r-i) + 127.9,
\end{equation}
    \item Having compared the gyrochronological models to our gyrochrone, we found that late K and early M dwarfs spin down as solid bodies between 2.7 and 4 Gyr of age. This behavior is broadly consistent with core-envelope decoupling models of stellar spin-down.
\end{itemize}

\begin{acknowledgments}
 RD would like to acknowledge the organizers of the Fifty Years of the Skumanich Relation conference for facilitating some enlightening discussions around these data, especially those with Jason Curtis.
 RD, JvS, and EG acknowledge support from NSF Astronomy \& Astrophysics grant AST-1817215.
 ARGS acknowledges the support from the FCT and FEDER/COMPETE2020 through work contract No. 2020.02480.CEECIND/CP1631/CT0001 and grants UIDP/04434/2020; PTDC/FIS-AST/30389/2017 \& POCI-01-0145-FEDER-030389.
 R.A.G.\ acknowledges the support from PLATO CNES grant.
 S.M.\ acknowledges support by the Spanish Ministry of Science and Innovation with the Ramon y Cajal fellowship number RYC-2015-17697 and the grant number PID2019-107187GB-I00.
 Based on observations obtained with MegaPrime/MegaCam, a joint project of CFHT and CEA/DAPNIA, at the Canada-France-Hawaii Telescope (CFHT) which is operated by the National Research Council (NRC) of Canada, the Institut National des Science de l'Univers of the Centre National de la Recherche Scientifique (CNRS) of France, and the University of Hawaii.
 The observations at the Canada-France-Hawaii Telescope were performed with care and respect from the summit of Maunakea which is a significant cultural and historic site.
 The Pan-STARRS1 Surveys (PS1) and the PS1 public science archive have been made possible through contributions by the Institute for Astronomy, the University of Hawaii, the Pan-STARRS Project Office, the Max-Planck Society and its participating institutes, the Max Planck Institute for Astronomy, Heidelberg and the Max Planck Institute for Extraterrestrial Physics, Garching, The Johns Hopkins University, Durham University, the University of Edinburgh, the Queen's University Belfast, the Harvard-Smithsonian Center for Astrophysics, the Las Cumbres Observatory Global Telescope Network Incorporated, the National Central University of Taiwan, the Space Telescope Science Institute, the National Aeronautics and Space Administration under Grant No. NNX08AR22G issued through the Planetary Science Division of the NASA Science Mission Directorate, the National Science Foundation Grant No. AST-1238877, the University of Maryland, Eotvos Lorand University (ELTE), the Los Alamos National Laboratory, and the Gordon and Betty Moore Foundation.
\end{acknowledgments}

\vspace{5mm}
\facilities{CFHT (MegaPrime), Gaia, PS1}

\software{astropy \citep{astropy2013, astropy2018},  
          photutils \citep{photutils},
          numpy \citep{numpy},
          scipy \citep{scipy},
          matplotlib \citep{matplotlib},
          hdbscan \citep{hdbscan},
          jupyter \citep{jupyter},
          astroquery \citep{astroquery}
          }

\appendix

\section{Example Light Curves and Periodograms} 
\label{extra_figures}

A light curve, phase-folded light curve, and periodogram are available for every star in Table \ref{tab:results}, included here are two examples. The full set is available in the online journal.

\figsetstart
\figsetnum{16}
\figsettitle{Light Curves and Periodograms}

\figsetend

\begin{figure}
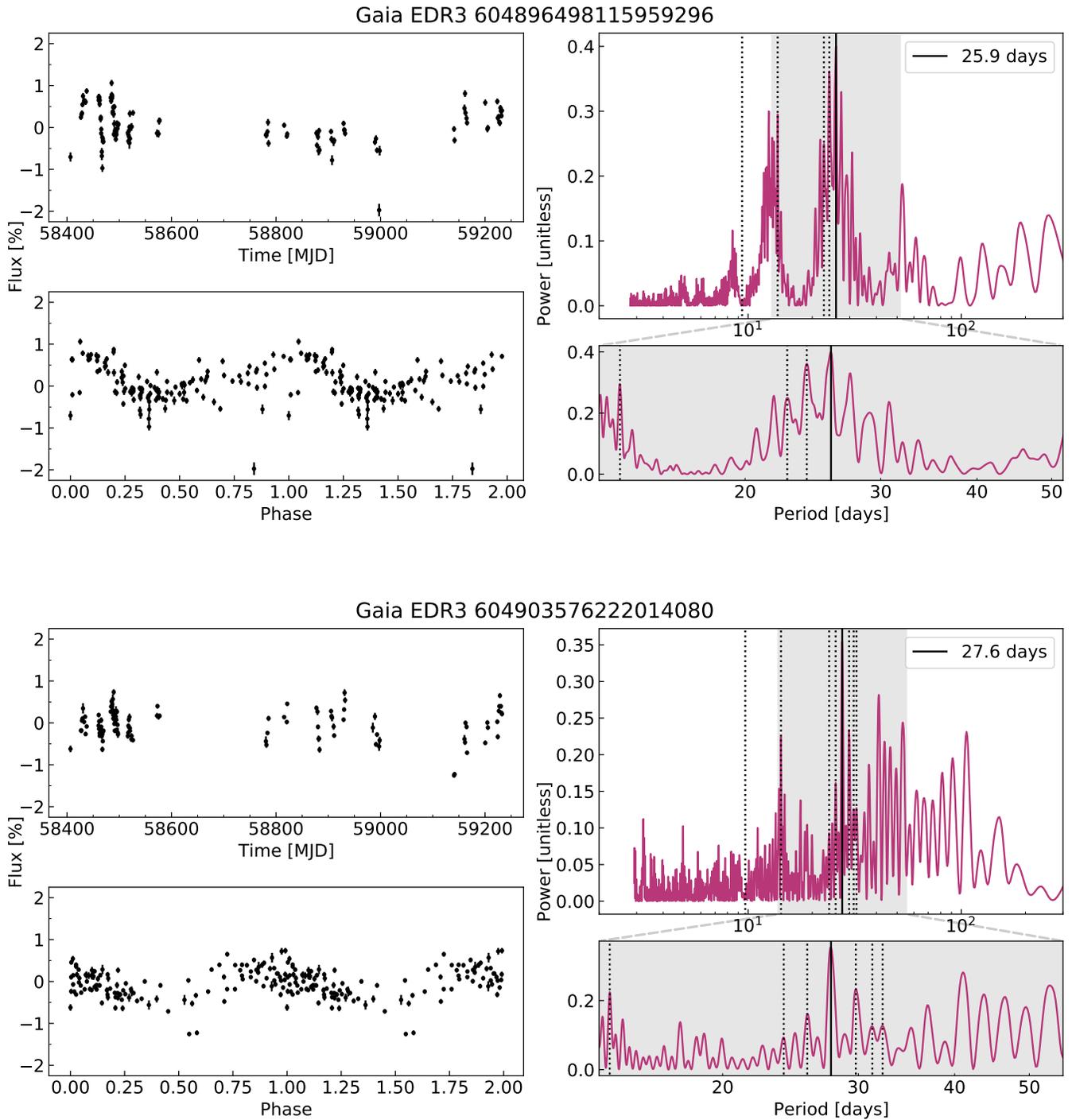

    \gridline{\fig{lc_example1.pdf}{1.0\textwidth}{}}
    \gridline{\fig{lc_example2.pdf}{1.0\textwidth}{}}
    \caption{Example light curves and periodograms for select cluster members. In the top left of each target's set is the light curve of the target. Bottom left is the phase-folded version, folded on the period of max power in the periodogram. On the right, the periodogram and a subset centered on the period of max power. Window peaks are denoted with vertical dotted lines. The complete figure set (594 images) is available in the online journal.}
    \label{fig:example_lcs}
\end{figure}

\bibliography{m67_prot}{}
\bibliographystyle{aasjournal}

\end{document}